\documentclass[twocolappendix]{emulateapj}
\slugcomment{Accepted by the Astrophysical Journal 2013 December 26; Published 2014 February 12}
\usepackage{amsmath}
\usepackage{amssymb}
\usepackage{graphicx,graphics}
\usepackage{rotating}
\usepackage{natbib}
\usepackage{apjfonts}
\usepackage{hyperref}
\hypersetup
{
    pdfauthor={Kokubo et al.},
    pdftitle={kokubo2014},
}
\shorttitle{Spectral Variability of SDSS quasars}
\shortauthors{Kokubo et al.}

\begin{document}

\title{STATISTICAL PROPERTIES OF MULTI-EPOCH SPECTRAL VARIABILITY OF SDSS Stripe~82 QUASARS}

\author{Mitsuru Kokubo\altaffilmark{1}}
\author{Tomoki Morokuma\altaffilmark{1}}
\author{Takeo Minezaki\altaffilmark{1}}
\author{Mamoru Doi\altaffilmark{1,2,3}}
\author{\\Toshihiro Kawaguchi\altaffilmark{4}}
\author{Hiroaki Sameshima\altaffilmark{5}}
\author{Shintaro Koshida\altaffilmark{6}}
\affil{${}^{1}$\ Institute of Astronomy, School of Science, 
the University of Tokyo, 2-21-1 Osawa, Mitaka,Tokyo 181-0015, Japan; \href{mailto:mkokubo@ioa.s.u-tokyo.ac.jp}{mkokubo@ioa.s.u-tokyo.ac.jp}\\
${}^{2}$\ Kavli Institute for the Physics and Mathematics of the Universe (WPI),
Todai Institutes for Advanced Study, \\
The University of Tokyo, 
5-1-5 Kashiwanoha, Kashiwa, Chiba 277-8583, Japan\\
${}^{3}$\ Research Center for the Early Universe, 
the University of Tokyo, 
7-3-1 Hongo, Bunkyo-ku, Tokyo 113-0033, Japan\\
${}^{4}$\ Department of Physics and Information Science,
Yamaguchi University,
Yamaguchi, Yamaguchi 753-8512, Japan\\
${}^{5}$\ Institute of Space and Astronautical Science, 
Japan Aerospace Exploration Agency, 
3-1-1 Yoshinodai, Sagamihara, Kanagawa 252-5210, Japan\\
${}^{6}$\ Center of Astro Engineering and Department of Electrical Engineering, 
Pontificia Universidad Catolica de Chile, 
Av. Vicu\~na Mackenna 4860, Santiago, Chile}

\begin{abstract}
We investigate the UV$-$optical (longward of Ly$\alpha$ 1216\AA) spectral variability of nearly 9000 quasars ($0<z<4$) using multi-epoch photometric data within the SDSS Stripe~82 region.
The regression slope in the flux$-$flux space of a quasar light curve directly measures the color of the flux difference spectrum, then the spectral shape of the flux difference spectra can be derived by taking a careful look at the redshift dependence of the regression slopes.
First, we confirm that the observed quasar spectrum becomes bluer when the quasar becomes brighter.
We infer the spectral index of the composite difference spectrum as $\alpha_{\nu}^{\text{dif}}\sim +1/3$ (in the form of $f_{\nu}\propto \nu^{\alpha_{\nu}}$), which is significantly bluer than that of the composite spectrum $\alpha_{\nu}^{\text{com}}\sim -0.5$.
We also show that the continuum variability cannot be explained by the accretion disk models with varying mass accretion rate.
Second, we examine the effects of broad emission line variability on the color-redshift space.
The variability of the ''Small Blue Bump" is extensively discussed. We show that the low-ionization lines of \ion{Mg}{2} and \ion{Fe}{2} are less variable compared to Balmer emission lines and high-ionization lines, and the Balmer continuum is the dominant variable source around $\sim 3000$\AA.
These results are compared with previous studies, and the physical mechanisms of the variability of the continuum and emission lines are discussed.
\end{abstract}


\section{Introduction}
\label{sec:1}
The flux variability of quasars, or active galactic nuclei (AGNs), have been observed on timescales ranging from hours to decades. UV$-$optical emissions are thought to arise from an accretion disk surrounding a central supermassive black hole.
UV$-$optical variability has been considered a powerful tool for examining the nature of AGN central engine \citep[][and references therein]{van04,kou12}.

Spectral variability (i.e., wavelength dependence of the flux variation) provides important clues for underlying physics.
One of the well-known features of AGN UV$-$optical variability is that the color tends to become bluer when it gets brighter \citep[e.g.,][]{mao93,pal94,giv99,van04,sak11,sch12,zuo12}, which is used to constrain the accretion disk models in several previous works \citep[e.g.,][and references therein]{sch12}.
This bluer-when-brighter trend is also important for studies of the emission line formation, as a change of the spectral energy distribution of the ionizing continuum has significant effects on the physical state of the emission line region \citep[e.g.,][]{kor04}.
However, as noted in  \cite{sak10}, the ``observed'' bluer-when-brighter trend admits a dual interpretation: (1) the variable component, which is probably the accretion disk emission itself, becomes brighter and gets bluer  \citep[e.g.,][]{wam90,giv99,web00,van04,wil05}, and (2) the variable component of constant color becomes brighter and increasingly dominant over the non-variable components of red color, which is mainly composed of host galaxy flux  \citep[e.g.,][]{cho81,win92,win97,pal96,sug06}.
That is, the ``observed'' bluer-when-brighter trend does not directly imply AGN intrinsic spectral hardening.
Moreover, we have to be careful of the effect of the broad emission lines (BELs) when analyzing the broad-band photometric variability, because they have large amount of flux and are variable themselves.
 \cite{sak10} investigated the color variability of 11 nearby Seyfert galaxies, estimating the effect of non-variable components, and concluded that interpretation (2) is valid for the AGN intrinsic continuum emission in the optical region, which kept the constant spectral shape during the flux variation.
The same result was obtained by  \cite{woo07} and  \cite{wal09}.
On the other hand,  \cite{sak11} did the same kind of analyses as \cite{sak10} for 10 Sloan Digital Sky Survey (SDSS) high redshift quasars and concluded that AGN intrinsic spectra actually became steeper as it got brighter in the UV region, which indicates that interpretation (1) is valid in the UV region.
\cite{wil05} used two-epoch spectral data for $\sim$ 300 SDSS quasars
and concluded that the composite flux difference spectrum is steeper than the composite spectrum in the UV range but it has the same spectral index in the optical range.
Note that the effects of the contamination from non-variable components, which are undoubtedly existing at least in part in the composite spectrum \citep[e.g.,][]{gli06,van06} are not evaluated in \cite{wil05}.
Thus, a consensus about the properties of the color variability of AGNs has not been obtained yet  \citep[see e.g.,][]{bia12,zha13}.

The difficulties of variability studies have led to the uncertainties of the AGN variability model.
A number of models have been developed attempting to explain AGN variability such as X-ray reprocessing  \citep{kro91}, instabilities in the accretion disk  \citep{kaw98,dex11}, gravitational microlensing  \citep{haw93}, star collisions  \citep{tor00}, and multiple supernovae or starbursts near the nucleus  \citep{tel92}.
Recently, several authors have claimed that the AGN variability is due to the changes in accretion rate  \citep{per06,Li08,sak11,zuo12,gu13}, which seems to explain the larger variability in the shorter wavelengths quantitatively.
\cite{sak11} fitted the standard accretion disk model \citep{sha73} with a varying accretion rate for the light curves of 10 SDSS quasars and concluded that this model could correctly explain the color variability.
\cite{gu13} also obtained the same conclusion for eighteen steep spectrum radio quasars.
However,  \cite{sch12} investigated the color variability of $\sim 9000$ SDSS Stripe~82 quasars and concluded that the color variability of quasars cannot be described by the several accretion disk models  \citep{dav07} with varying accretion rate, as the color of quasars becomes bluer than the model predictions.
\cite{trv02} also pointed out that the changes of accretion rate are insufficient to
explain the amount of spectral variation.

In the present work, we focus on the model of \cite{per06}, which assumes that the AGN variability is caused by small changes in the mass accretion rate in the standard thermal accretion disk  \citep{sha73} and predict the flux difference spectrum between one epoch to another.
We note that although we cannot directly observe the intrinsic AGN spectra or intrinsic color, we can directly observe the color of the variable portion in AGN spectra (i.e., the flux difference spectrum).
This can be achieved by using the ``flux$-$flux correlation method'' described in Section~\ref{sec:3}, which relies on the fact that UV$-$optical, two-band flux to flux plots of AGN broad-band light curves show strong linear correlation  \citep[e.g.,][and references therein]{cho81,sak11,lir11}.
As mentioned earlier, the flux variability occurs not only in the continuum emission, but also in the BELs.
In general, the quasar variability (obtained by broad-band photometry) is dominated by continuum variability \citep[intrinsic Baldwin effect, e.g.,][]{wil05,shi07}, but we are probably able to see some amount of variable emission line components in it.
We can also examine the effects of the BEL variability on the broad band photometry by taking a careful look at our result, which has not been well studied in previous works.

In this paper, we investigate the years-time-scale UV$-$optical spectral variability of large quasar samples in SDSS Stripe~82 region, attempting to derive some insight to the physical mechanisms of AGN variability.
Our standpoint is summarized as below:
\begin{itemize}
\item Using the flux$-$flux correlation method, we derive the color of the flux difference spectrum for each quasar and for each band pair model-independently.
Then, we investigate the rest-frame wavelength dependence of the quasar spectral variability by taking a careful look at the redshift dependence of the color of the flux difference spectrum, with the assumption that there is no redshift-dependence in the rest-frame spectral variability \citep[e.g.,][]{sch12,zuo12}.
The derived flux difference spectrum can be directly compared with the AGN UV$-$optical variability model presented in  \cite{per06}.
\item We do not assume any models or estimates of non-variable components such as host galaxy flux or narrow line emissions in the present work.
This means that the color variability trend (bluer-when-brighter or achromatic variation) referred to in this paper is actually the ``observed'' color variability trend, not the AGN intrinsic variation in spectral shape.
However, as shown by  \cite{sak11}  \citep[see also][]{cro09,kra13}, host galaxy flux and narrow emission lines in SDSS bandpasses are probably negligible for intermediate and high redshift quasars. Thus the ``observed'' color variability trend, particularly in rest-frame UV wavelengths, can be considered as the intrinsic variation in AGN continuum spectral shape, although the effects of the variability of BELs should be taken into consideration.
\end{itemize}

In Section~\ref{sec:2} we describe a database of SDSS Stripe~82 multi-epoch five-band light curves from the SDSS data, and we introduce the flux$-$flux correlation method  \citep{cho81} in Section~\ref{sec:3}.
In Section~\ref{sec:4}, we show the linear regression slopes in flux$-$flux space for quasar light curves, which correspond to the color of the flux difference spectrum, as a function of redshift.
Then, we compare our results with previous works and the standard thin-disk model.
We discussed the emission line variability in Section~\ref{sec:5}. The variability of the Small Blue Bump (SBB) spectral region is discussed separately in Section~\ref{sec:6}.
Finally, we summarize our conclusions in Section~\ref{sec:7}.
The adopted models for emission lines and intergalactic medium (IGM) attenuation used in Section~\ref{sec:5} are described in Appendix~\ref{sec:A}, and the results of the regression intercept, which are not concerned with the main results in this paper, are interpreted and shown in Appendix~\ref{sec:B}.

\section{Data}
\label{sec:2}
We use a database of SDSS Stripe~82 multi-epoch five-band light curves for spectroscopically confirmed quasars presented by \cite{mac12}.
Their Southern sample catalog contains recalibrated $\sim $10 yr light curves for 9258 quasars in the SDSS Data Release 7 \citep[DR7; ][]{aba09}.
SDSS magnitude is converted to flux unit.
The SDSS system is nearly the AB system (zero-point flux density $=$ 3631 Jy), but the photometric zero-points are slightly off the AB standard.
We apply the correction for the zero-point offset between the SDSS system and the AB system following the recommendations in the SDSS instructions\footnote{\href{http://www.sdss.org/dr7/algorithms/fluxcal.html}{http://www.sdss.org/dr7/algorithms/fluxcal.html}}: 
$u_{\text{AB}}=u_{\text{SDSS}}-0.04$, $g_{\text{AB}}=g_{\text{SDSS}}$, $r_{\text{AB}}=r_{\text{SDSS}}$, $i_{\text{AB}}=i_{\text{SDSS}}$ and $z_{\text{AB}}=z_{\text{SDSS}}+0.02$ .
In this paper, ``Flux'' means $f_{\nu}$ unit (i.e., Jansky unit), but we also use $f_{\lambda}$ unit when we refer to the spectral shape.
These two are related as $f_{\nu}=\lambda ^2 f_{\lambda}/c$, and power-law indices are related as $\alpha_{\nu}=-2-\alpha_{\lambda}$, where $f_{\nu} \propto \nu ^{\alpha_{\nu}}$ and $f_{\lambda} \propto \lambda ^{\alpha_{\lambda}}$.

\begin{figure}[tbp]
 \begin{center}
  \includegraphics[clip, width=3.4in]{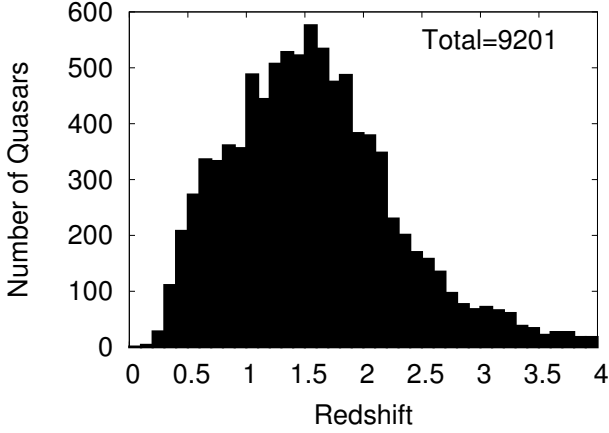}
  \vspace{-0.5cm}
 \end{center}
 \caption{Redshift distribution ($0< z <4$) of the SDSS Stripe~82 quasar sample.}
 \label{fig:redshift_distribution}
\end{figure}

We cross-match the Southern sample light curve catalog with a catalog of quasar properties from SDSS DR7  \citep{she11}(the maximum search radius is 2 arcsec).
We use the improved redshift from \cite{hew10} as the redshift of each quasar.
In the present paper, we use a sample of quasars within the redshift range
from 0 to 4, in which there are 9201 quasars.
The redshift distribution of the sample ($0< z <4$) is shown in Figure~\ref{fig:redshift_distribution}.

Galactic absorption is corrected using $A_u$[mag] tabulated in the database of \cite{mac12} for each quasar, where $A_g$, $A_r$, $A_i$, $A_z$ $=$ 0.736, 0.534, 0.405, 0.287 $\times$ $A_u$ \citep{sch98,mac12}.

\section{The flux-flux correlation method}
\label{sec:3}
In this section, we introduce the flux$-$flux correlation method and discuss the advantages of the use of this method.

\subsection{The Color of the Flux Difference Spectrum in Flux-Flux Space}
\label{sec:3_1}

\begin{figure}[tbp]
 \begin{center}
  \includegraphics[clip, width=3.4in]{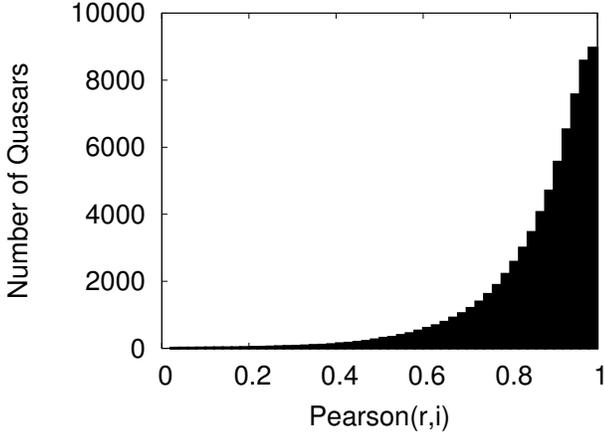}
  \vspace{-0.5cm}
 \end{center}
 \caption{Cumulative distribution of Pearson correlation coefficient in flux$-$flux space ($x$-axis $=$ $r$-band flux, and $y$-axis $=$ $i$-band flux) for light curves of our sample ($0 < z <4$) observed more than 20 epochs (see Section~\ref{sec:3_5} for more details of our sample).}
 \label{fig:pearson_r}
\end{figure}

\begin{figure}[tbp]
 \begin{center}
\includegraphics[clip, width=3.4in]{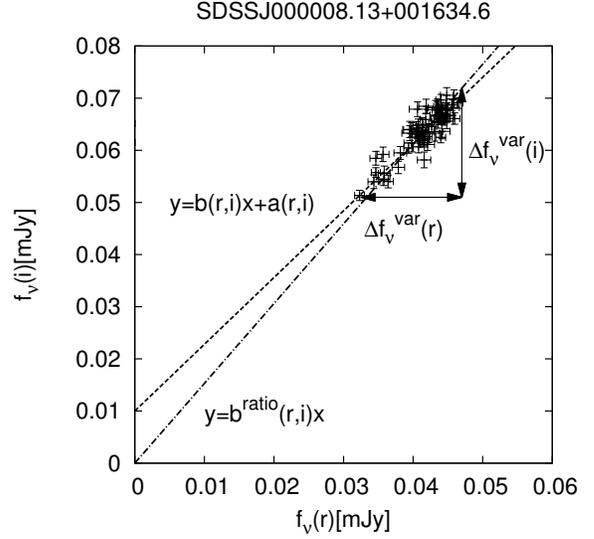}
\vspace{-0.7cm}
 \end{center}
\caption{Illustration for the flux$-$flux correlation method.
The linear regression line is $y=1.28x+0.01$ [mJy] (i.e., $b(r,i)=\Delta f^{\text{var}}_{\nu}(i)/\Delta f^{\text{var}}_{\nu}(r)=1.28$.
The time-averaged color is $b^{\text{ratio}}(r,i)=\bar{f}_{\nu}(i)/\bar{f}_{\nu}(r)=1.53$, so the flux difference spectrum is bluer than the time-averaged spectrum in this object.}
\label{fig:ff}
\end{figure}

We introduce flux$-$flux correlation method \citep{cho81,win92,lyu93,hag97} as an alternative to magnitude$-$magnitude or magnitude$-$color correlation analyses.
\cite{cho81} first noticed that the quasars' UV$-$optical fluxes have a linear correlation in flux$-$flux space, and several authors have since confirmed it  \citep[e.g.,][]{win97,gla04,sak11,lir11,gu13}.
In Figure~\ref{fig:pearson_r}, we show the Pearson correlation coefficient of the $f_{\nu}(r)-f_{\nu}(i)$ plot for our sample ($0 < z <4$) observed more than 20 epochs (see Section~\ref{sec:3_5} for more details of our sample).
In terms of the Pearson correlation coefficient, almost all of the quasars have strong positive flux$-$flux correlation.
The strong correlation for a quasar light curve means that the flux difference 
spectrum, which is defined as the flux difference of the quasar spectra of any 
of the two observational epochs, keeps nearly constant shape for at least 
several years.

We express a linear regression line ($y=b\times x+a$) in flux$-$flux space ($f_{\nu}(s)-f_{\nu}(l)$ space) as
\begin{eqnarray}
f_{\nu}(l)&=&b(s,l)\times f_{\nu}(s)+a(s,l)\label{linear}\\ 
(s&=&u,g,r,i ; \ l=g,r,i,z) \nonumber
\end{eqnarray}
where ``$s$'' (``short'' wavelength, $x$-axis in flux$-$flux space), and ``$l$'' (``long'' wavelength, $y$-axis in flux$-$flux space) indicate two of the five photometric bands, whose average wavelengths are $\lambda(s)$ $<$ $\lambda(l)$ (where $\lambda(u)$,     
 $\lambda(g)$, $\lambda(r)$, $\lambda(i)$, $\lambda(z)$ $=$ 3551, 4686, 6165, 7481, 8931\AA )\footnote{\href{http://www.sdss.org/dr7/instruments/imager/index.html}{http://www.sdss.org/dr7/instruments/imager/}}.
Figure~\ref{fig:ff} shows an illustration of our method.
When we fit a regression line in flux$-$flux space,
the regression slope $b(s,l)$ and the intercept $a(s,l)$ are expressed as
\begin{eqnarray}
b(s,l)&=&\frac{\Delta f_{\nu}^{\text{var}}(l)}{\Delta f_{\nu}^{\text{var}}(s)} \label{b_def}\\
a(s,l)&=&\bar{f}_{\nu}(l)-b(s,l)\bar{f}_{\nu}(s)
\label{a_def}
\end{eqnarray}
where $\Delta f_{\nu}^{\text{var}}$ is the varying broad-band flux range (flux difference) during the observations, and $\bar{f}_{\nu}$ is the baseline flux given as the mean flux of light curves for each bandpass.
The effects of contaminations of baseline flux ($=\bar{f}_{\nu}$, e.g., host galaxy flux and time-averaged fluxes of emission lines) do not affect to the regression slope (they only affect to the regression intercepts), so the regression slope in the flux$-$flux space directly measures the ratio of the two broad-band fluxes of the flux difference spectrum of each quasar.

Given the flux difference spectrum $f_{\nu}^{\text{dif}}$ for a quasar, the regression slope in flux$-$flux space is expressed as  \citep{ric01}, 
\begin{equation}
b(s,l)=\frac{\int_{}^{}f_{\nu}^{\text{dif}}S_{\nu}(l)d\log \nu}{\int_{}^{}f_{\nu}^{\text{dif}}S_{\nu}(s)d\log \nu}\times \left( \frac{\int_{}^{}S_{\nu}(l)d\log \nu}{\int_{}^{}S_{\nu}(s)d\log \nu}\right)^{-1}
\label{b_cal}
\end{equation}
where $S_{\nu}$ is the throughput of the SDSS photometric system in each bandpass  \citep{doi10}.

From Equation~(\ref{b_cal}), the color of the flux difference spectrum
$m_s-m_l|_{\text{dif}}$ (e.g., $r-i|_{\text{dif}}$) in units of magnitude is expressed by $b(s,l)$ as,
\begin{equation}
m_s-m_l|_{\text{dif}}=+2.5\log (b(s,l)).
\label{b_color}
\end{equation}

This means that the bluer color indicates a smaller value of Equation~(\ref{b_color}), and then a smaller value of $b(s,l)$.
In this way, the color of the flux difference spectrum is related directly to the observable $b(s,l)$.
In the present work, we refer $b(s,l)$ simply as ``the color of the flux difference spectrum'' and this can be converted to the conventional definition of color through Equation~(\ref{b_color}).

If we assume a power-law difference spectrum ($f_{\nu}^{\text{dif}}\propto \nu ^{\alpha_{\nu}^{\text{dif}}}$),  we obtain (by Equation~(\ref{b_cal}))
\begin{equation}
b(s,l) \sim \left( \frac{\nu(l)}{\nu(s)} \right)^{\alpha_{\nu}^{\text{dif}}}=\left( \frac{\lambda(s)}{\lambda(l)} \right)^{\alpha_{\nu}^{\text{dif}}}
\label{approximate_b}
\end{equation}
where $\nu(s)$ and $\nu(l)$ indicate average frequencies for the SDSS photometric bands.
Because we define $b(s,l)$ as $\lambda(l)>\lambda(s)$,
\begin{equation}
\begin{cases}
    b(s,l)>1 & (\text{if} \ \alpha_{\nu}^{\text{dif}}<0) \\
    b(s,l)<1 & (\text{if} \ \alpha_{\nu}^{\text{dif}}>0).
  \end{cases}
\label{spectral_variability}
\end{equation}
When the flux difference spectra of quasars have power-law shape, the color of the flux difference spectra has no dependence on the quasar redshift because the wavelength-shift by cosmic expansion is canceled out in Equation~(\ref{approximate_b}).
That is, $b(s,l)$ with a power-law difference spectrum is constant as a function of redshift if $\alpha_{\nu}^{\text{dif}}$ is fixed.
Although the color of the difference spectrum is actually contaminated by BEL variability, 
we can infer $\alpha_{\nu}^{\text{dif}}$ by investigating the mean redshift dependence of $b(s,l)$ for sample quasars.

\subsection{The Color of Quasars in Flux-Flux Space}
\label{sec:3_2}

In flux$-$flux space, the ``observed'' color corresponds to the slope of a straight line that passes through the origin of the coordinates and an observed point at an epoch.
The ``observed'' color is time-dependent if $a(s,l)$ (the regression intercept in flux$-$flux space) is non-zero \citep[][and see Appendix B]{sak10,sak11}.

We define the $b^{\text{ratio}}(s,l)$ by taking the time-average of light curves in any combination of five photometric bands (``baseline flux''  $\bar{f}_{\nu}(j)$, $j=u$, $g$, $r$, $i$, $z$) and taking the ratio as (see Figure~\ref{fig:ff})
\begin{equation}
b^{\text{ratio}}(s,l)\equiv \frac{\bar{f}_{\nu}(l)}{\bar{f}_{\nu}(s)}.
\label{b_ratio}
\end{equation}
Equation~(\ref{b_color}), which is the (time-averaged) color $m_1-m_2$ of the quasar spectrum (e.g., $r-i$) is expressed by $b^{\text{ratio}}(s,l)$ as
\begin{equation}
m_s-m_l=+2.5\log (b^{\text{ratio}}(s,l)).
\label{b_ratio_mag}
\end{equation}
We refer to $b^{\text{ratio}}(s,l)$ simply as ``the color of the quasar spectrum'' in this paper, and you can always convert it to the conventional definition of color through Equation~(\ref{b_ratio_mag}).

$b^{\text{ratio}}(s,l)$ is expressed for a quasar with a spectrum $f_{\nu}$ as
\begin{equation}
b^{\text{ratio}}(s,l)=\frac{\int_{}^{}f_{\nu}S_{\nu}(l)d\log \nu}{\int_{}^{}f_{\nu}S_{\nu}(s)d\log \nu}\times \left( \frac{\int_{}^{}S_{\nu}(l)d\log \nu}{\int_{}^{}S_{\nu}(s)d\log \nu}\right)^{-1}.
\label{b_ratio_cal}
\end{equation}
If we assume a power-law spectrum ($f_{\nu} \propto \nu ^{\alpha_{\nu}}$), we obtain
\begin{equation}
b^{\text{ratio}}(s,l) \sim \left( \frac{\nu(l)}{\nu(s)} \right)^{\alpha_{\nu}}=\left( \frac{\lambda(s)}{\lambda(l)} \right)^{\alpha_{\nu}}.
\label{approximate_b_ratio}
\end{equation}
Because we define $b^{\text{ratio}}(s,l)$ as $\lambda (l) > \lambda(s)$, 
\begin{equation}
\begin{cases}
    b^{\text{ratio}}(s,l)>1 & (\text{if} \ \alpha_{\nu}<0) \\
    b^{\text{ratio}}(s,l)<1 & (\text{if} \ \alpha_{\nu}>0).
  \end{cases}
\label{spectral}
\end{equation}

\subsection{The Color-Redshift Relation}
\label{sec:3_3}

\cite{ric01} investigated the causes of the observed features in the color as a function of redshift for 2625 SDSS quasars by comparing the color$-$redshift relation model including the effects of emission lines.
They confirmed that the  average of the observed color of quasars as a function of redshift
is well represented by a power-law continuum using $\alpha_{\nu}\sim -0.5$  \citep[similar to the result of ][]{van01}, 
and identified the effects of emission lines for the broad-band colors as deviations from power-law colors.

\begin{figure}[tbp]
 \begin{center}
  \includegraphics[clip, width=3.0in]{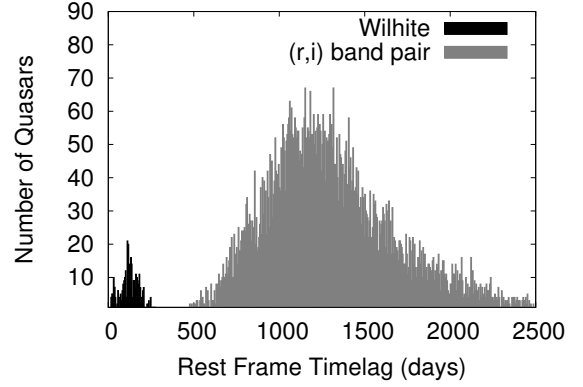}
  \vspace{-0.3cm}
 \end{center}
 \caption{Rest-frame time-lag histograms binned for five days.
 The left histogram (black, median = 123 days) indicates the number of quasars with two-epoch spectra used in \cite{wil05} (time-lag is defined as the difference of the observational epochs).
The right histogram (gray, median $\sim$ 1200 days) is the rest-frame time-lag distribution of $r$ and $i$-band light curves, which we used in the present work (time-lag is defined as the difference of first and last observational epochs).
See the details of our sample definition in Section~\ref{sec:3_5}.}
 \label{fig:ri_timescale}
\end{figure}

Because we are able to obtain the color of the flux difference spectrum for quasars using the flux$-$flux correlation method, the same kind of analyses as  \cite{ric01} for the flux difference spectrum can be applied.
Currently, there are only a few studies related to the flux difference spectrum  \citep{wil05,bia12,guo13}.
Among these, we refer to that of \cite{wil05} for the quasar composite flux difference spectrum in this paper.
They constructed the composite residual (composite flux differences spectrum) from 315 quasars observed twice by SDSS spectroscopy.
Figure~\ref{fig:ri_timescale} shows the histogram of the rest-frame time-lag between the two spectroscopic observations in Wilhite's sample, with a histogram of the time span of the photometric observations in our sample.
The median rest-frame time-lag of the Wilhite's sample is 123 days, which is about one-tenth of ours (median $\sim$ 1200 days).
Aside from the difference in the rest-frame time-lag, we can compare our results with the (redshifted) composite flux difference spectrum by  \cite{wil05} using Equation~(\ref{b_cal}) on the color$-$redshift space.
Note that  \cite{wil05} mainly focused on the ratio spectrum (the ratio of the composite difference spectrum to the composite spectrum), so they did not apply the correction of Galactic extinction to their composite difference spectrum.
We discuss the effect introduced by this non-correction in Section~\ref{sec:4_2}.

\begin{figure}[tbp]
 \begin{center}
  \includegraphics[clip, width=3.0in]{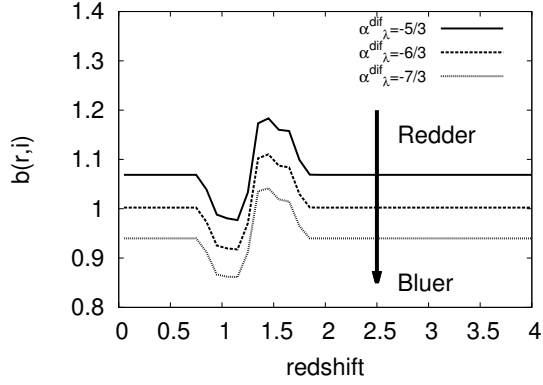}
  \vspace{-0.3cm}
 \end{center}
 \caption{Illustration of an emission line contamination effect on the $b(r,i)-$redshift plane.
We can treat $b(s,l)$ and $b^{\text{ratio}}(s,l)$ in the same way, so here we show the case of $b(s,l)$.
Curves are calculated by Equation~(\ref{b_cal}) combined with Equation~(\ref{line_var}), for $\alpha^{\text{dif}}_{\lambda}=-5/3,-6/3,-7/3$.
The power-law spectrum with definite spectral index means constant color at all redshift (Equation~(\ref{approximate_b})).
At $z$ $\sim$ 1.1, the emission line component of Equation~(\ref{line_var}) enters the $r$-band, and the color becomes bluer than the power-law color.
At $z$ $\sim$ 1.5, the emission line component leaves the $r$-band and enters the $i$-band, and the color becomes redder than the power-law color.}
 \label{line_contami_ri}
\end{figure}

Here, we show the illustration of the effect of an emission line contamination 
for $b(s,l)$ and $b^{\text{ratio}}(s,l)$ on the redshift dependence  \citep[see also][]{ric01}.
We can treat $b(s,l)$ and $b^{\text{ratio}}(s,l)$ in the same way, so here we show the case of $b(s,l)$.
We consider the case of $s=r$ and $l=i$ (i.e., $b(r,i)$) for example.
When an emission line is redshifted and enters $r$-band, the $r$-band flux becomes larger, and $b(r,i)$ becomes smaller (bluer) than the case of a single power-law flux difference spectrum.
At higher $z$, when the emission line enters the $i$-band, $b(r,i)$ becomes larger (redder).
This behavior is shown in Figure~\ref{line_contami_ri}, assuming the continuum and line variability as
\begin{eqnarray}
f^{\text{dif}}_{\lambda} &=&  f^{\text{dif}}_{\lambda}(\text{continuum})+f^{\text{dif}}_{\lambda}(\text{line}) \nonumber \\
&=&\left(\frac{\lambda}{3000\text{\AA}}\right)^{\alpha^{\text{dif}}_{\lambda}}+\frac{1}{2}\exp\left(-\frac{(\lambda-\lambda_{\text{rest}})^2}{2\sigma_{\lambda}^2}\right) \label{line_var}
\end{eqnarray}
where $\sigma_{\lambda}$ is the emission line width (here $\sigma_{\lambda}=40$\AA), and $\lambda_{\text{rest}}$ is the central wavelength of the line (here $\lambda_{\text{rest}}=3000\text{\AA}$). 
$\alpha_{\lambda}^{\text{dif}}$ is as shown in Figure~\ref{line_contami_ri}.
The different power-law indices $\alpha^{\text{dif}}_{\lambda}$ lead to the different constant levels of $b(s,l)$ following Equation~(\ref{approximate_b}).

\subsection{Advantages of the Flux-Flux Correlation Method}
\label{sec:3_4}

 \cite{sch12} pointed out that 
previous analyses usually focused on the color variability of the quasar in color$-$magnitude space and suffered from covariance between the color and magnitude uncertainties.
They avoided these error correlations by analyzing the color 
variability in magnitude$-$magnitude space, and concluded that the color variability has the bluer-when-brighter trend.

Although magnitude$-$magnitude correlation method can avoid error correlations, it still has another problem.
\cite{sch12} fitted a straight line for each 
quasar's magnitude$-$magnitude plot, and used the best fit slope as the indicator of color variability.
Note that regression slopes on magnitude$-$magnitude space highly depend on the contaminations of baseline flux (e.g., host galaxy flux and time-averaged fluxes of emission lines) \citep[e.g.,][]{woo07,haw03}.
The regression slope in magnitude$-$magnitude space can be expressed as
\begin{eqnarray}
&&b^{\text{mag}}(s,l) \approx \nonumber\\
&&\frac{\log [\left( \bar{f}_{\nu}(l)+\Delta f_{\nu}^{\text{var}}(l)/2\right)/\left(\bar{f}_{\nu}(l)-\Delta f_{\nu}^{\text{var}}(l)/2 \right)]}{\log [\left( \bar{f}_{\nu}(s)+\Delta f_{\nu}^{\text{var}}(s)/2\right)/\left(\bar{f}_{\nu}(s)-\Delta f_{\nu}^{\text{var}}(s)/2 \right)]}\nonumber
\end{eqnarray}
where the notation is the same as Equation~(\ref{b_def}) and (\ref{a_def}).
Unlike in flux$-$flux space (Equation~(\ref{b_def})), the regression slope in magnitude-magnitude space is dependent not only on the flux difference $\Delta f^{\text{var}}_{\nu}$, but also on the time-averaged flux levels $\bar{f}_{\nu}$ (``baseline flux'', Equation~(\ref{b_ratio})).
Thus, $b^{\text{mag}}(s,l)$ only tells whether the ``observed'' color variability is bluer-when-brighter ($b^{\text{mag}}<1$), redder-when-brighter ($b^{\text{mag}}>1$) or achromatic ($b^{\text{mag}}=1$).
To derive the flux difference spectrum from $b^{\text{mag}}$, one needs to employ some models or estimate the contamination $\bar{f}_{\nu}$.
On the other hand, the flux$-$flux correlation method enables us not only to avoid error correlations, but also to derive the flux difference spectrum model independently (Equation~(\ref{b_def})).

Another statistical method to derive the flux difference spectrum exists.
\cite{meu11} investigated the rest-frame wavelength dependence of the quasar variability using the structure functions $V^{\text{mag}}$ of five SDSS photometric bands as a function of rest-frame wavelength.
However, the structure function analyses, which are done in units of magnitude, are not appropriate to study the spectral variability or the flux difference spectrum because the structure function can be expressed as
\begin{eqnarray}
V^{\text{mag}} &\approx& [\Delta m]^2 \nonumber\\
&\approx& \left[ -2.5\log \left( \frac{\bar{f}_{\nu}(s)+\Delta f^{\text{var}}_{\nu}(s)/2}{\bar{f}_{\nu}(s)-\Delta f^{\text{var}}_{\nu}(s)/2} \right) \right]^2.\nonumber
\end{eqnarray}
Then again, this is dependent not only on the flux difference $\Delta f^{\text{var}}_{\nu}$, but also on the time-averaged flux levels $\bar{f}_{\nu}$.

In summary:
\begin{enumerate}
\item The flux$-$flux correlation method can avoid the error correlations that exist in magnitude$-$color or flux$-$color correlation method as was pointed out by  \cite{sch12}.
\item The regression slope $b(s,l)$ derived by the flux$-$flux correlation of a quasar multi-band light curve is not affected by baseline fluxes (e.g., host galaxy flux or time-averaged fluxes of emission lines).
\item Regression slopes represent the color of the flux difference spectrum for each quasar, so we are able to infer the (mean) flux difference spectrum using redshift dependence of the derived regression slopes.
\end{enumerate}

\subsection{Linear Regression Method and Sample Selection}
\label{sec:3_5}

Linear regression for the data that have intrinsic scatter ($\sigma _{\text{int}}$) and error bars on both axes (e.g., flux$-$flux space linear regression) is complex and there are several different methods \citep[see][and references therein]{par12,fei12,cap13}.
We adopt the {\tt MPFITEXY} IDL routine \citep{wil10} to fit a linear relation to the flux$-$flux correlation for each quasar. According to  \cite{par12}, the FITEXY method generally provides the least-biased result compared to the other methods.
The {\tt MPFITEXY} routine depends on the {\tt MPFIT} package  \citep{mar09}.
This routine is able to cope with intrinsic scatter, which is automatically adjusted to ensure $\chi ^2$/(degrees or freedom) $\sim $1  \citep[see][for more detail and justifications]{tre02,nov06,par12}.

We fit a straight line in flux$-$flux space for each band pair light curve, excluding the data points with a ``Bad observations'' flag \citep{mac12} or fainter than limiting magnitudes (22.0, 22.2, 22.2, 21.3, and 20.5mag for $u$, $g$, $r$, $i$, and $z$-band, respectively).\footnote{\href{http://www.sdss.org/dr7/}{http://www.sdss.org/dr7/}} 
We focused on the properties of long-term and multi-epoch-averaged variability, so we exclude quasars that have less than 20 photometric epochs.
Linear relation $y=b(x-x_0)+a'$ is assumed in the fitting procedure, where $x_0$ is taken to be the average of $x$ values (i.e., $\bar{f}_{\nu}(s)$).
The reference value $x_0$ is necessary to minimize the uncertainty in the estimate of $a'$ and the correlation between $a'$ and $b$  \citep{tre02}.
Regression intercept $a$ is calculated as $a = a' - b x_0$.
Linear regression for several quasars did not converge because of the low signal-to-noise ratio (S/N).
Thus, we eliminated them from our sample.
Furthermore, we confirmed that all of the regression slopes with a negative value were the results of a bad fitting (because of the small variability or low S/N) and we also eliminated them from our sample.
In this paper we examine the modal redshift trends, so these eliminations had little effect on our analyses.

\begin{figure}[tbp]
 \begin{center}
  \includegraphics[clip, width=3.0in]{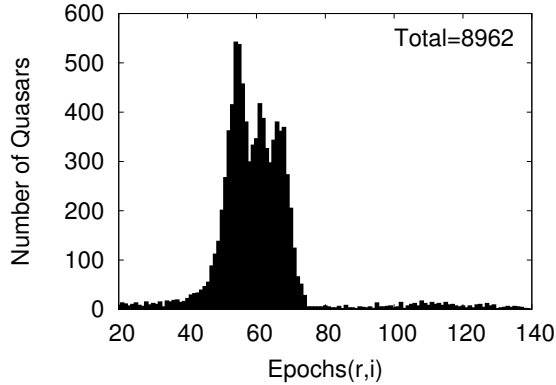}
  \vspace{-0.3cm}
 \end{center}
 \caption{Histogram of the number of observational epochs of the ($r$,$i$) band pair light curves for our quasar sample.
 Observations with a ``Bad observations'' flag  \citep{mac12} are excluded.}
 \label{fig:ri_epochs}
\end{figure}

As an example, we show the histogram of the number of observational epochs of the ($r$,$i$) band pair light curves without a ``Bad observations'' flag  \citep{mac12} for our quasar sample in Figure~\ref{fig:ri_epochs}.
Resulting sample sizes for each band pair are shown in Table \ref{sample_size}.
Redshift distribution is not significantly modified from Figure~\ref{fig:redshift_distribution} by our sample selection.

\begin{deluxetable}{cc}
\tablecaption{Sample Size ($0<z<4$)}
\tablehead{\colhead{($s,l$)} & \colhead{Samples}}
\startdata
($u,g$) & 8,537 \\
($u,r$) & 8,523 \\
($u,i$) & 8,495 \\
($u,z$) & 8,213 \\
($g,r$) & 8.996 \\
($g,i$) & 8,949 \\
($g,z$) & 8,745 \\
($r,i$) & 8,962 \\
($r,z$) & 8,770 \\
($i,z$) & 8,761
\enddata
\label{sample_size}
\end{deluxetable}

\begin{figure*}[!t]
\centerline{
\hspace{-0.1cm}
\includegraphics[clip, width=7.5in]{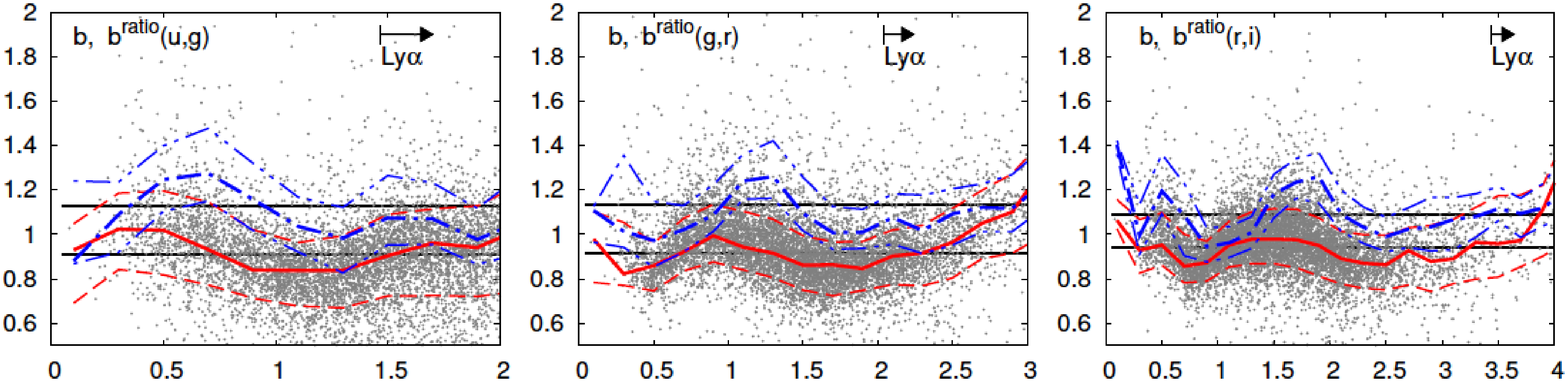}
}
\centerline{
\hspace{-0.0cm}
\includegraphics[clip, width=7.5in]{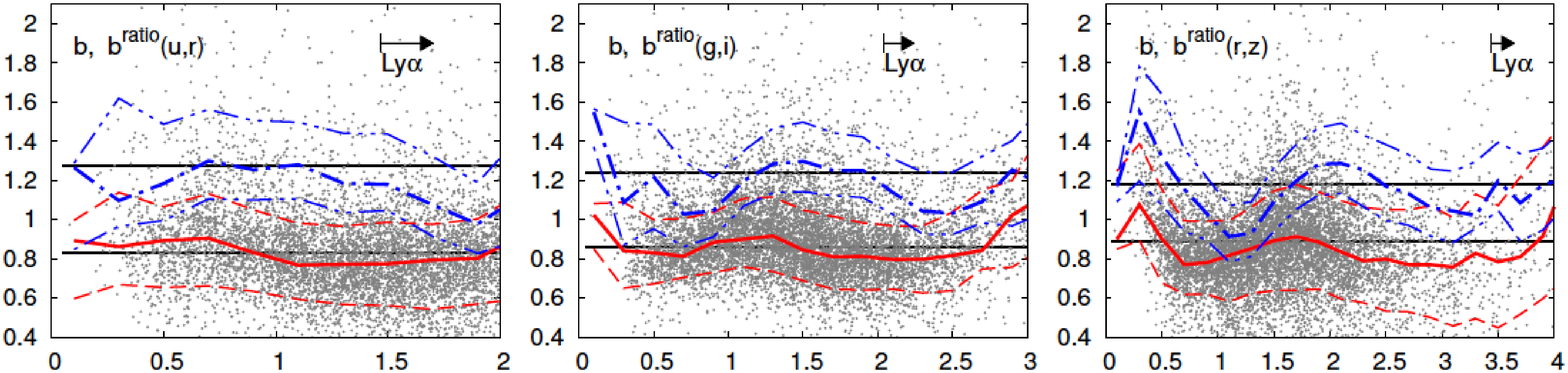}
}
\centerline{
\hspace{-0.05cm}
\includegraphics[clip, width=7.55in]{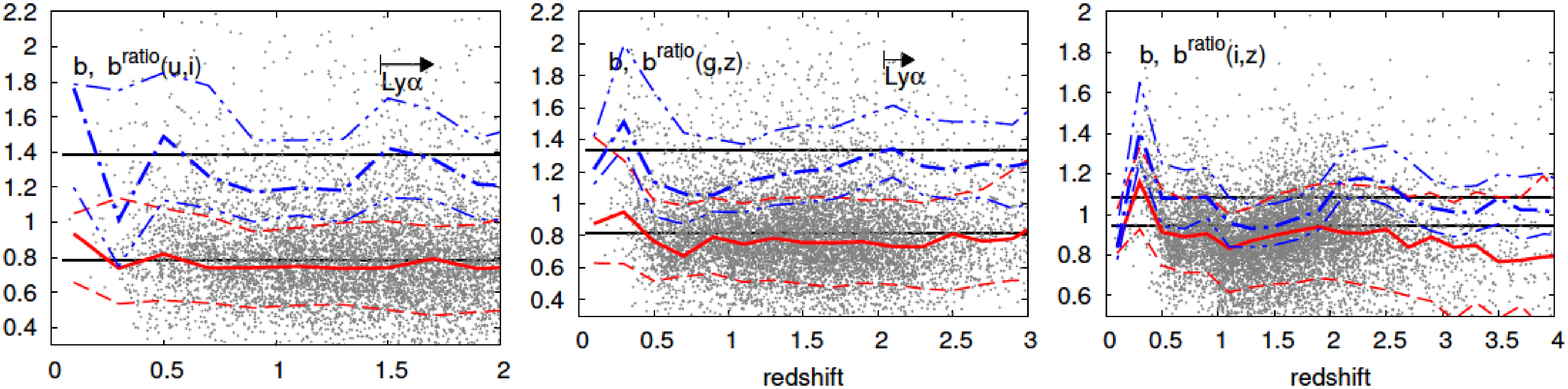}
}
\leftline{
\hspace{-0.67cm}
\includegraphics[clip, width=2.57in]{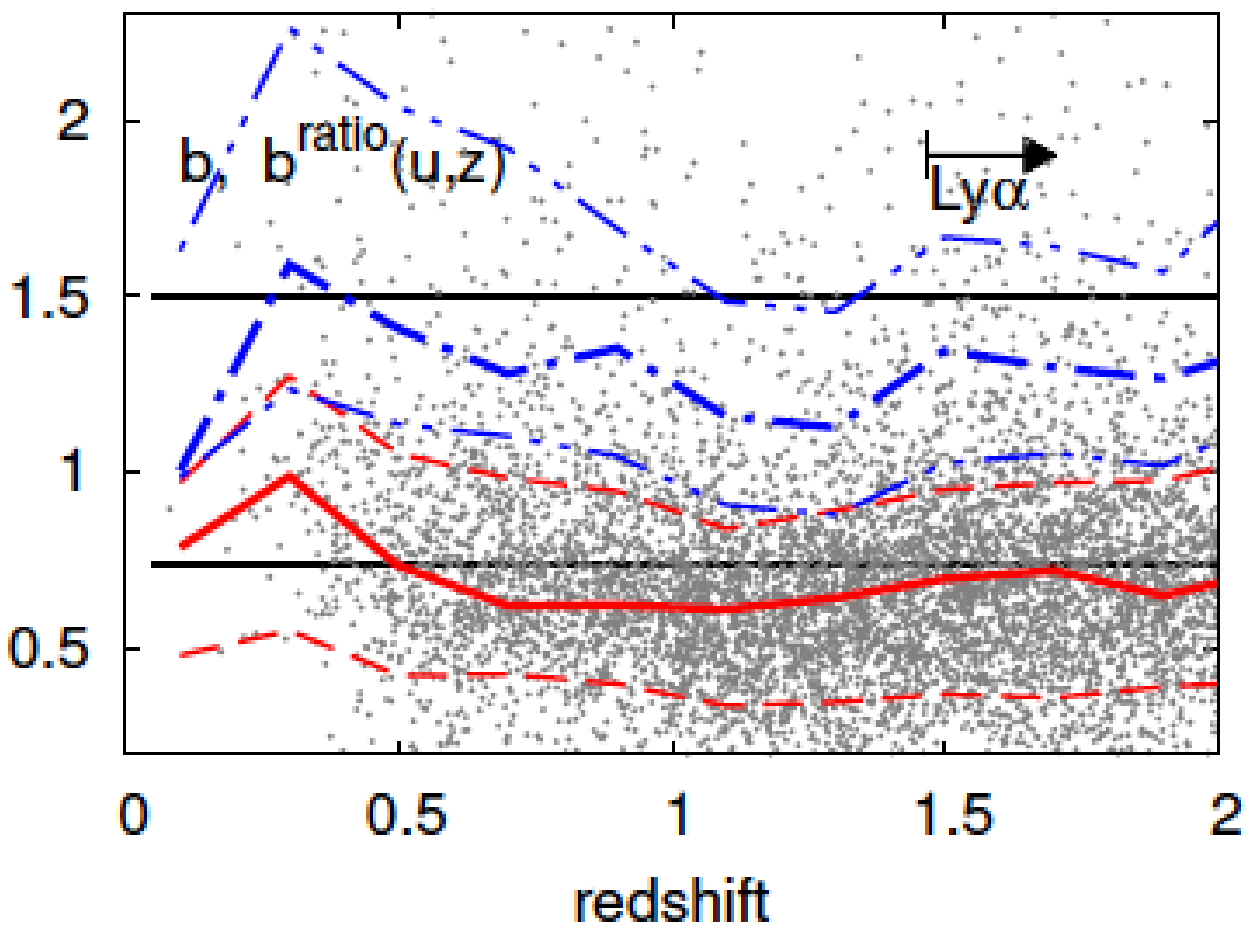}
}
\caption{Regression slopes $b(s,l)$ as a function of redshift (dots).
Solid curves with dashed outer quartiles (colored red in the electronic edition) indicate the mode and the 1$\sigma$ (68\%) range of the slopes for each redshift bins ($\Delta z$ $=$ 0.2), and dash-dotted curves with double-dot-dashed outer quartiles (colored blue in the electronic edition) are the mode of the ratio of fluxes $b^{\text{ratio}}(s,l)$ defined by Equation~(\ref{b_ratio}).
Two thin horizontal lines (upper and bottom) indicate the case of power-law spectrum models, $\alpha _{\nu}=$ $-0.44$ and $+1/3$ ($\alpha _{\lambda}=-1.56$, $-7/3$), respectively.
$b^{\text{ratio}}(s,l)-$redshift relation (i.e., the color of the time-averaged observed spectrum as a function of redshift) can be represented by $\alpha ^{\text{com}}_{\nu}=-0.44$ \citep{van01}.
On the other hand, $b(s,l)-$redshift relation (i.e., the color of the difference spectrum as a function of redshift) can be represented by $\alpha ^{\text{dif}}_{\nu}=+1/3$.}
\label{fig:b_ugriz}
\end{figure*}

\begin{figure*}[!t]
\centerline{
\hspace{-0.0cm}
\includegraphics[clip, width=7.5in]{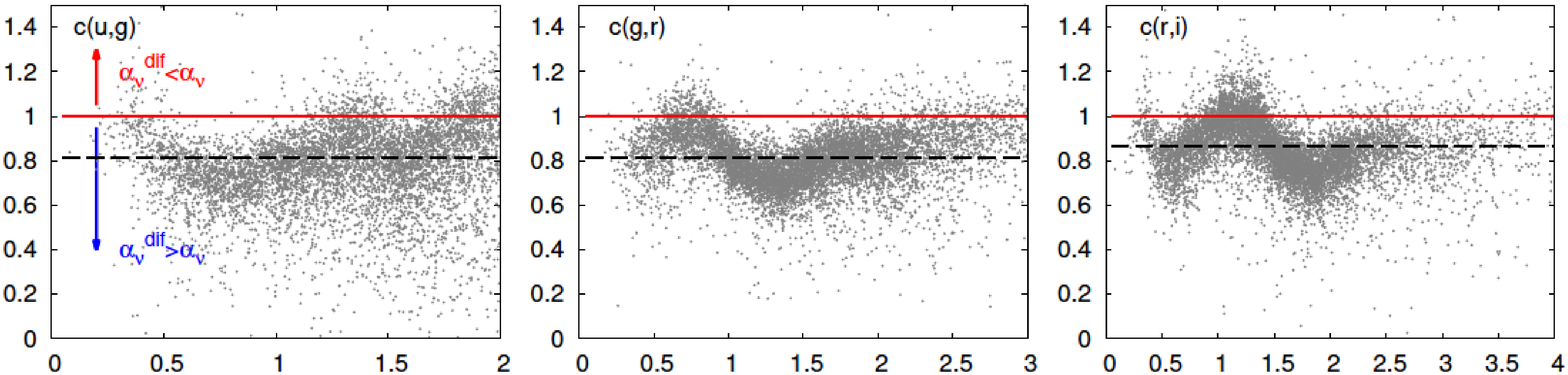}
}
\centerline{
\hspace{-0.0cm}
\includegraphics[clip, width=7.5in]{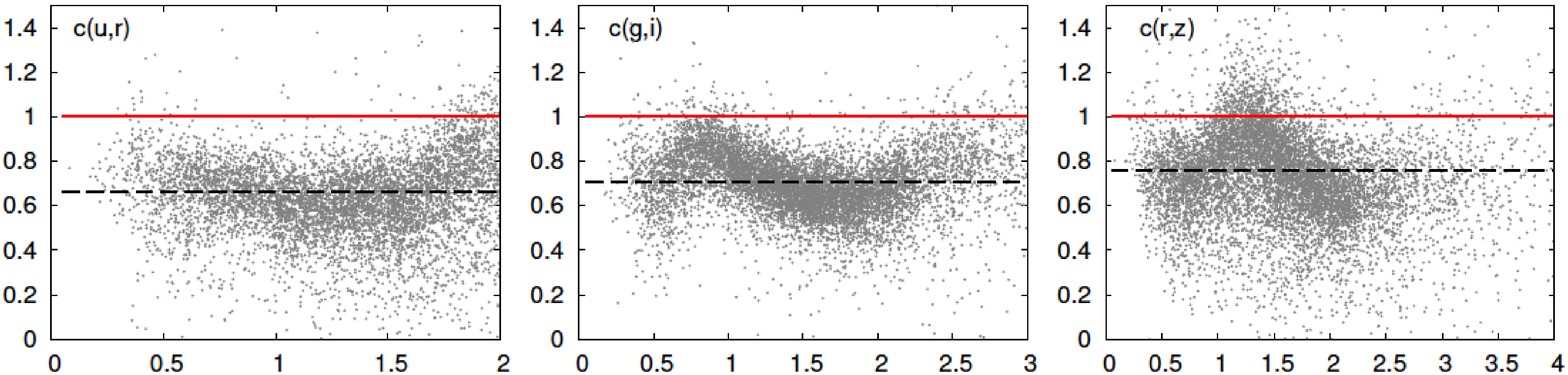}
}
\centerline{
\hspace{-0.0cm}
\includegraphics[clip, width=7.5in]{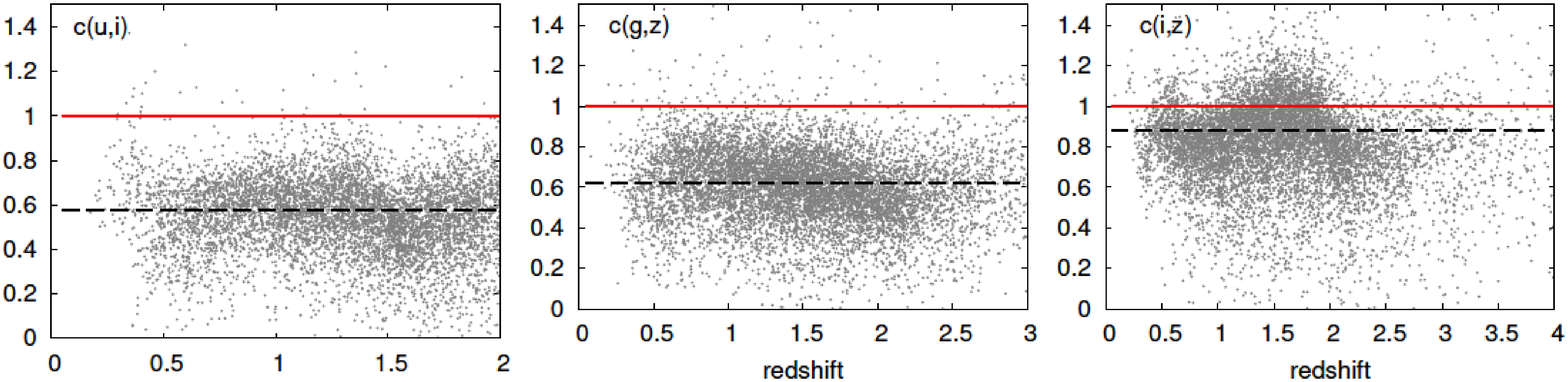}
}
\leftline{
\hspace{-0.6cm}
\includegraphics[clip, width=2.5in]{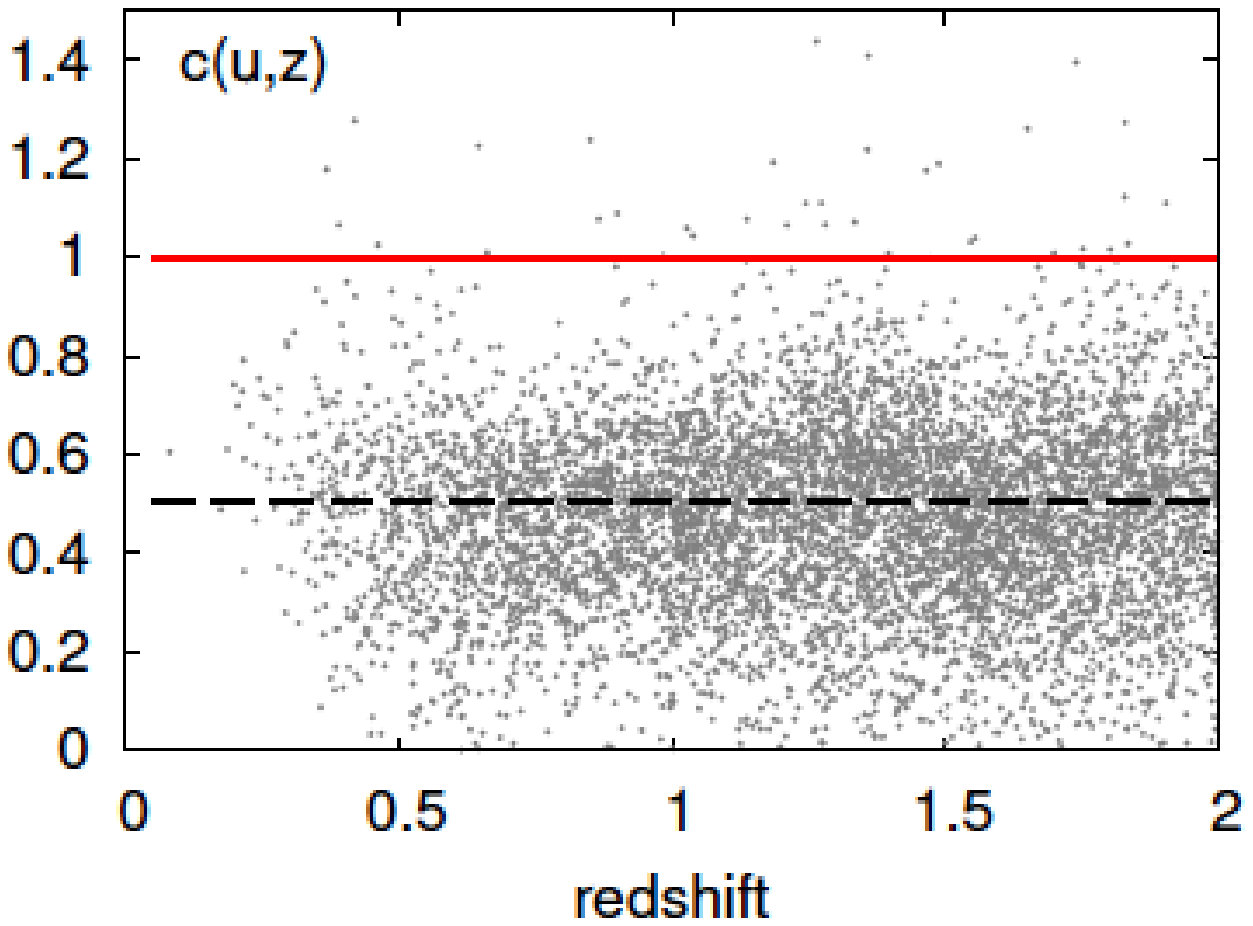}
}
\caption{$c(s,l)$ $\equiv {b(s,l)}/{b^{\text{ratio}}(s,l)}$ (Equation~(\ref{c_definition})) as a function of redshift.
The dashed horizontal lines indicate $\Delta \alpha_{\nu}=\alpha_{\nu}^{\text{dif}}-\alpha_{\nu} =-\alpha_{\lambda}^{\text{dif}}+\alpha_{\lambda}= +7/3 - 1.56 \sim 0.77$ (difference of the spectral indices shown in Figure~\ref{fig:b_ugriz}), and the solid lines (colored red in the electronic edition) indicate $\Delta \alpha_{\nu}=0.0$.
$c(s,l)$ $>$ 1 implies $\alpha_{\nu}^{\text{dif}}<\alpha_{\nu}$ (corresponds to redder-when-brighter trend in observed color), and $c(s,l)$ $<$ 1 implies $\alpha_{\nu}^{\text{dif}}>\alpha_{\nu}$ (corresponds to bluer-when-brighter trend in observed color).}
\label{fig:b_ugriz_ratio}
\end{figure*}

\section{Continuum Variability}
\label{sec:4}

In this section, we show linear regression slopes $b(s,l)$ derived  in flux$-$flux space for quasar light curves, which correspond to the color of the flux difference spectrum as discussed in the previous section.
The wide redshift range for our quasar samples and large separation of effective wavelengths of the SDSS filters enable us to investigate the continuum variability from 1216\AA\  to about 6000\AA \ in the rest-frame wavelength.
We first check the consistency between our result and the previous spectroscopic result of the flux difference spectrum \citep{wil05}.
Then, we adopt a model for the continuum flux variation in which the variability is caused by changes in mass accretion rate in the standard accretion disk \citep{per06} as a working hypothesis, and compare it with observed continuum variability in $b(s,l)-$redshift space.

\subsection{Redshift Dependence of the Color of the Flux Difference Spectrum}
\label{sec:4_1}

Figure~\ref{fig:b_ugriz} shows the regression slopes $b(s,l)$ as a function of redshift for each quasar (dots).
The curves in Figure~\ref{fig:b_ugriz} are:
\begin{itemize}
\item Solid and dashed curves: The modal color ($b(s,l)$) of the flux difference spectrum (solid) with $1\sigma$ outer quartiles (dashed), derived by linear regression analyses in flux$-$flux space (i.e., mode and $1\sigma$ outer quartiles of the dots in the figure).
\item Dash-dotted and double-dot-dashed curves: The modal color ($b^{\text{ratio}}(s,l)$) of the time-averaged spectrum (dash-dotted) with $1\sigma$ outer quartiles (double-dot-dashed), given by Equation~(\ref{b_ratio}).
\end{itemize}
Modal value is defined following  \cite{hop04}, and the bin size is taken to be 0.2 in redshift.

At high redshift, $u$, $g$, and $r$-band fluxes are affected by Ly$\alpha $ forests or damped Ly$\alpha$ systems, which absorb UV flux and decrease $u$, $g$, and $r$-band variability (i.e., regression slopes become larger at higher redshifts).
This makes it difficult to estimate intrinsic quasar spectrum and spectral variability.
The shortest transmission wavelengths for $u$, $g$, and $r$-band are 3000\AA ,\ 3700\AA ,\ and 5400\AA ,\ respectively, so the Ly$\alpha$ emission line (1216\AA) enters in these bands at $z=1.47, 2.04, 3.44$.
The arrows with the label ``Ly$\alpha$'' in Figure~\ref{fig:b_ugriz} indicate the redshift ranges in which Ly$\alpha $ forests change the observed color.
In the later sections, we focus on redshift range below $z=1.47$, $2.04$, and $3.44$ for band pairs containing $u$, $g$, and $r$-band, respectively, for clarifying the discussion about intrinsic spectral variability.

The dash-dotted curves ($b^{\text{ratio}}(s,l)$) in Figure~\ref{fig:b_ugriz}  are generally larger than 1.
This is consistent with the result of the geometric mean composite spectrum for SDSS quasars presented by  \cite{van01}; the spectral index of the composite spectrum $\alpha^{\text{com}}_{\nu}\sim -0.44 < 0$ ($\alpha^{\text{com}}_{\lambda}\sim -1.56$) indicates $b(s,l)>1$ (Equation~(\ref{spectral})).
The upper thin solid horizontal line in each panel of Figure~\ref{fig:b_ugriz} indicates the power-law spectrum model with $\alpha_{\nu}$ $=$ $-0.44$ calculated by Equation~(\ref{b_cal}).

On the other hand, the solid curves ($b(s,l)$) in Figure~\ref{fig:b_ugriz} are generally less than 1 and it implies $\alpha_{\nu}^{\text{dif}}>0$ (Equation~(\ref{spectral_variability})).
The composite flux difference spectrum presented by  \cite{wil05} was reported to have the spectral index $\alpha_{\nu}^{\text{dif}} = 0$, and it implies $b(s,l) = 1$ (Equation~(\ref{approximate_b})).
However, the solid curves in Figure~\ref{fig:b_ugriz} are less than 1 within the $1\sigma$ ranges (dashed curves).
We attempt to employ the spectral index $\alpha_{\nu}^{\text{dif}}=+1/3$ that is suggested by  \cite{sak10} (the lower thin, solid horizontal line in each panel of Figure~\ref{fig:b_ugriz}).
This spectral index seems to reproduce the mean value of the $b(s,l)$ for all the band pairs well.
This inconsistency between the result of  \cite{wil05} and ours is discussed in Section~\ref{sec:4_2}.

$b^{\text{ratio}}(s,l)>b(s,l)$ is the general trend for quasar variability as shown in Figure~\ref{fig:b_ugriz} (see also Appendix B).
This indicates that the quasar composite flux difference spectrum is flatter (bluer) than the composite spectrum.
In other words, quasar UV$-$optical spectra tend to become flatter (bluer) when in brighter phase.
To clarify the ``observed'' bluer-when-brighter trend for individual quasars, we show the ratio of $b(s,l)$ to $b^{\text{ratio}}(s,l)$ given as 
\begin{equation}
c(s,l) \equiv \frac{b(s,l)}{b^{\text{ratio}}(s,l)} \sim \left( \frac{\lambda(s)}{\lambda(l)} \right)^{\Delta \alpha_{\nu}}
\label{c_definition}
\end{equation}
as a function of redshift in Figure~\ref{fig:b_ugriz_ratio}, where $\Delta \alpha_{\nu}=\alpha^{\text{dif}}_{\nu}-\alpha_{\nu}$ (approximate value is from Equation~(\ref{approximate_b}) and (\ref{approximate_b_ratio})).
By definition, as in Equation~(\ref{spectral_variability}) and (\ref{spectral}),
\begin{equation}
\begin{cases}
    c(s,l)>1 & (\text{if} \ \alpha_{\nu}^{\text{dif}}<\alpha_{\nu}) \\
    c(s,l)<1 & (\text{if} \ \alpha_{\nu}^{\text{dif}}>\alpha_{\nu}).
  \end{cases}
\label{spectral_variability_to_spectral}
\end{equation}
Dashed horizontal lines in  Figure~\ref{fig:b_ugriz_ratio} indicate $\Delta \alpha_{\nu}=\alpha_{\nu}^{\text{dif}}-\alpha_{\nu} =-\alpha_{\lambda}^{\text{dif}}+\alpha_{\lambda}=7/3 -1.56 \sim 0.77$ (difference of the spectral indices shown in Figure~\ref{fig:b_ugriz}), and solid horizontal lines indicate the case of $\alpha_{\nu}^{\text{dif}}=\alpha_{\nu}$.
Figure~\ref{fig:b_ugriz_ratio} shows that almost all of quasars have $\alpha_{\nu}^{\text{dif}}>\alpha_{\nu}$.
This means the ``observed'' bluer-when-brighter spectral variability is a common trend in quasars.

Besides the continuum (power-law) variability, we are able to see several spectral features in Figure~\ref{fig:b_ugriz} and also in Figure~\ref{fig:b_ugriz_ratio}.
These features are mainly made up of the emission line contributions \citep[e.g.,][]{wil05}.
It is known that the emission line variability is smaller compared with the continuum variability \citep[intrinsic Baldwin effect, ][]{wil05}. As expected by the intrinsic Baldwin effect, the solid curves in Figure~\ref{fig:b_ugriz} show less features compared to the dash-dotted curves.
Nonetheless, we are still able to identify the coincidence and the difference of features (the redshift dependence) between the solid and dash-dotted curves.
The most apparent difference between them, which corresponds to the difference between the composite and composite difference spectrum, is the bump in intermediate redshifts (e.g., the redshift range $z\sim 1.0$-$1.5$ of $g-r$ band pair in Figure~\ref{fig:b_ugriz}).
This difference makes several sharp transitions on $c(s,l)$ as a function of redshift shown in Figure~\ref{fig:b_ugriz_ratio}.
The peaks of the bump of dash-dotted curves correspond to the spectral region of \ion{Mg}{2} (2800\AA) emission line and \ion{Fe}{2}  pseudo-continuum, and this peak is not seen in the solid curves.
It implies that the variability of these emissions are weak.
This point is discussed in Sections~\ref{sec:5} and \ref{sec:6}.

\subsection{Comparison with the Composite and Composite Difference Spectrum}
\label{sec:4_2}

\begin{figure}[!t]
\centerline{
\includegraphics[clip, width=2.95in]{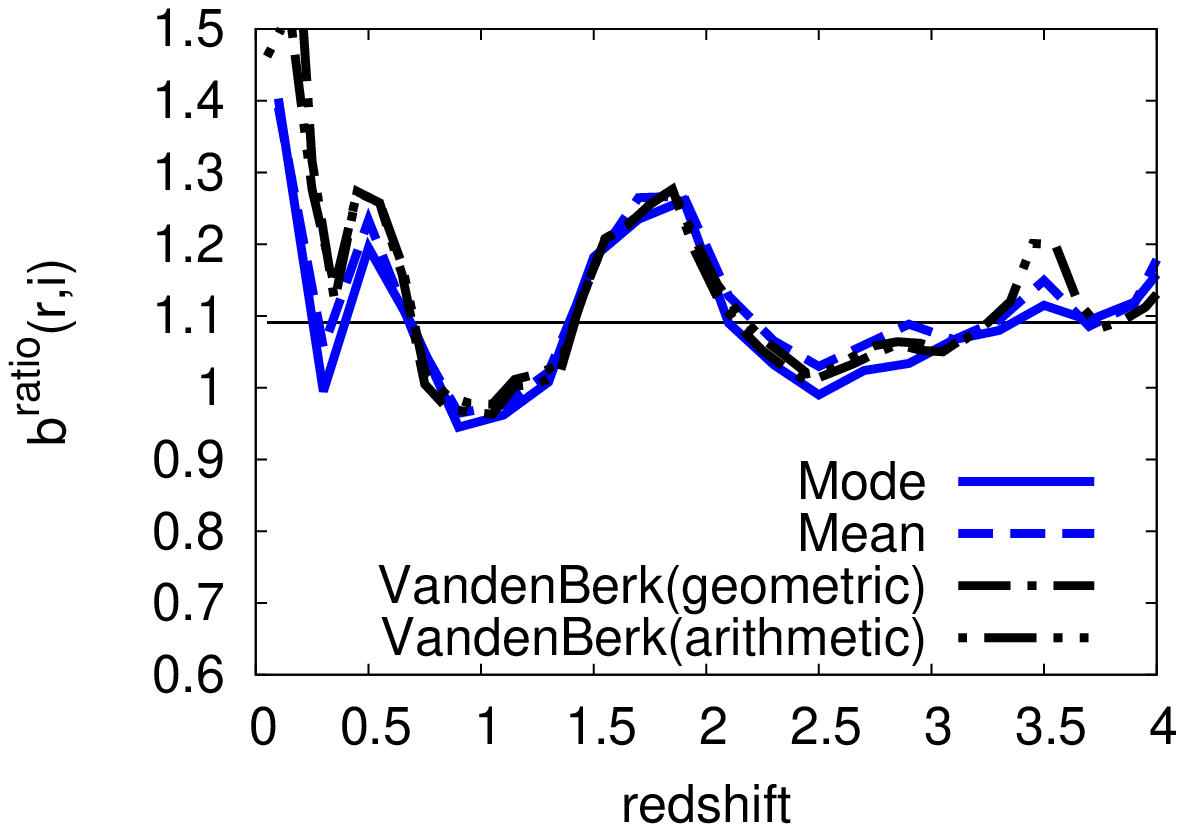}}
\centerline{
\includegraphics[clip, width=2.95in]{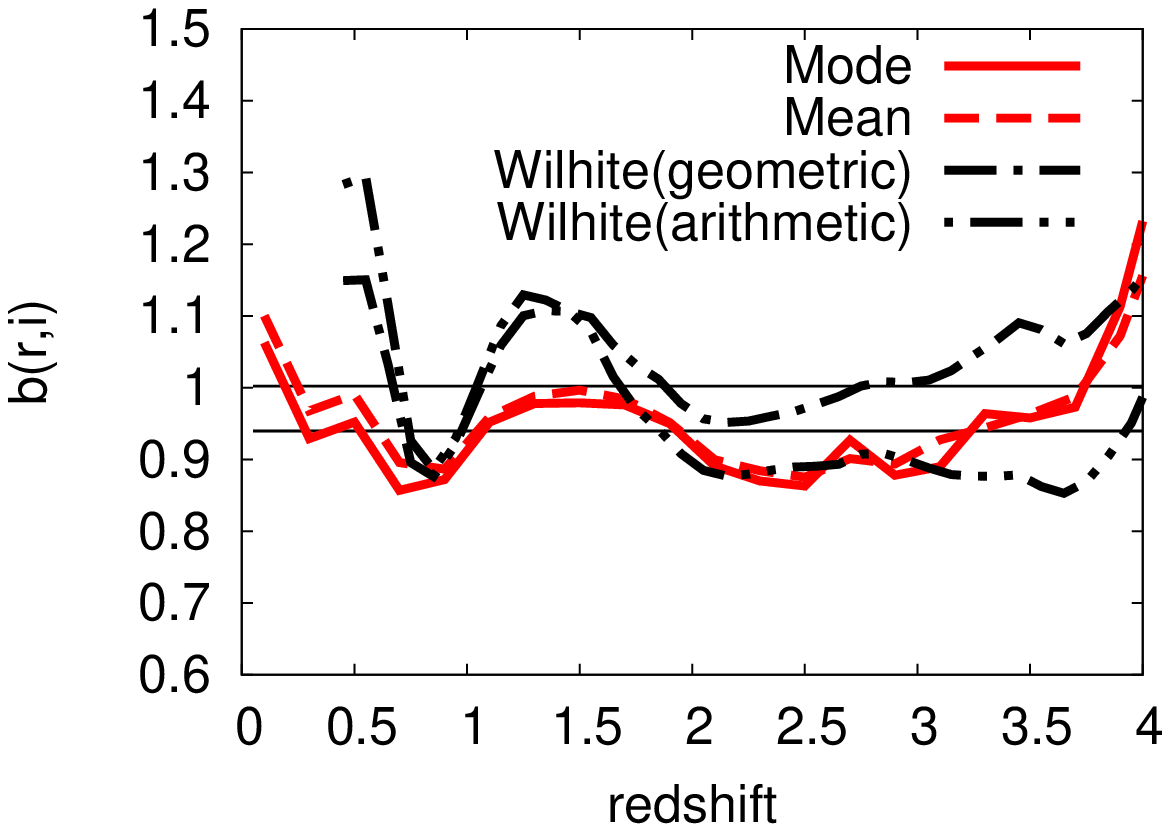}}
\centerline{
\includegraphics[clip, width=2.95in]{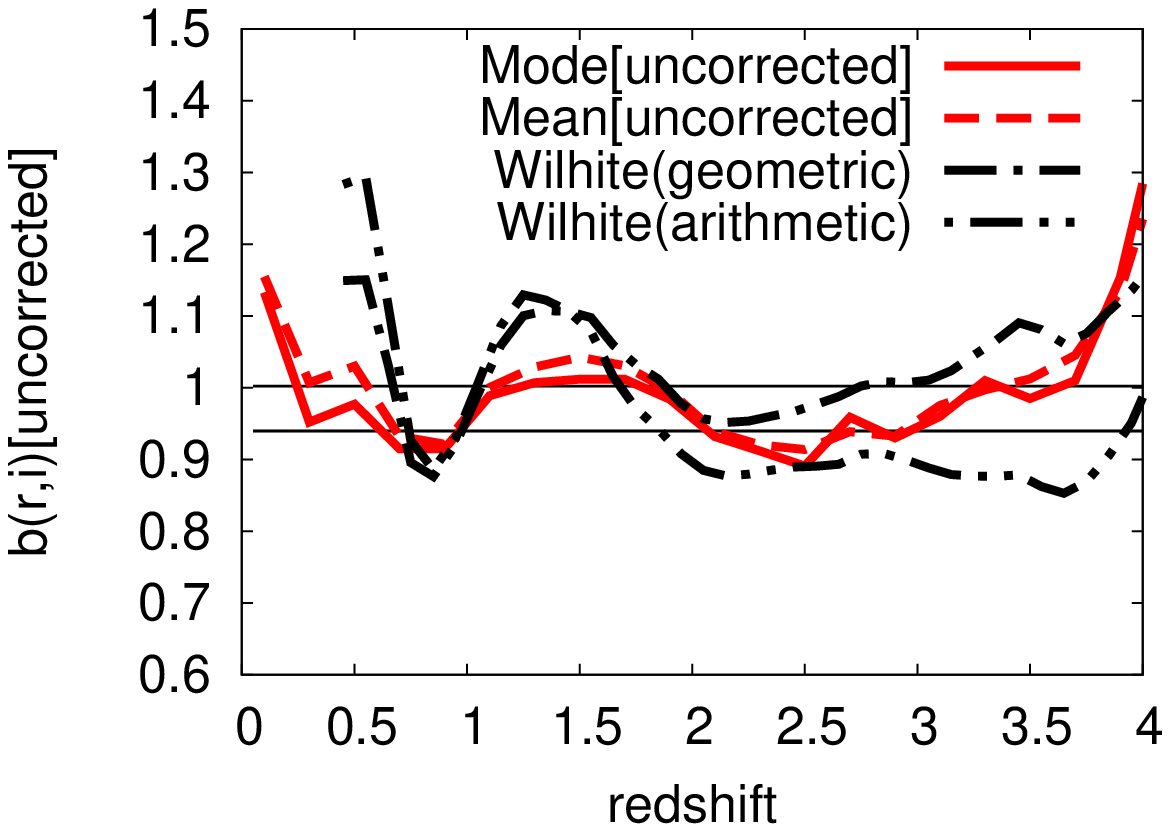}}
\caption{Top panel: the comparison of the modal and mean color of the time-averaged spectrum ($b^{\text{ratio}}(r,i)$) with the color of the redshifted geometric and arithmetic composite spectrum \citep{van01} calculated by Equation~(\ref{b_ratio_cal}).
The horizontal line is the color of a power-law spectrum with $\alpha_{\nu}=-0.44$ \citep{van01}.
Middle panel: the comparison of the modal and mean color of the difference spectrum ($b(r,i)$) with the color of the redshifted geometric and arithmetic composite difference spectrum \citep{wil05} calculated by Equation~(\ref{b_cal}).
The two horizontal lines are the color of power-law difference spectra with $\alpha_{\nu}^{\text{dif}}=$ 0 and +1/3 (upper and lower, respectively).
Bottom panel: the same plot as middle panel, but the modal and mean color of the difference spectrum ($b(r,i)$) is not corrected for Galactic extinction.}
\label{fig:comparison}
\end{figure}

There have been several statistical studies about spectral shape and spectral variability for quasars selected and spectroscopically confirmed within SDSS.
Here we show the consistency of these previous spectroscopic results with our photometric results.

Figure~\ref{fig:comparison} shows the comparison of our results (modal color as a function of redshift in Figure~\ref{fig:b_ugriz}) with the composite \citep{van01} and composite difference \citep{wil05} spectrum using Equation~(\ref{b_ratio_cal}) and (\ref{b_cal}).
In Figure~\ref{fig:comparison} we plot not only modal value, but also (weighted) mean value as a function of redshift to check the effects of the asymmetric color distribution, which has a red wing at a given redshift due to intrinsic dust reddening  \citep{ric03}.
We are able to see in Figure~\ref{fig:comparison} that the mean values are slightly larger (redder) than the modal value, but the difference is small.
This indicates that our results are not affected heavily by this method of taking average.

The composite spectra composed by taking arithmetic or geometric means  \citep{van01,wil05} are compared with our result through Equation~(\ref{b_cal}) and (\ref{b_ratio_cal}).
The geometric mean preserves the average power-law slope, insofar as quasar spectra can be accurately described by a power-law.
On the other hand, the arithmetic mean retains the relative strength of the non-power-law features, such as emission lines  \citep{van01}.

As shown in the top panel of Figure~\ref{fig:comparison}, $b^{\text{ratio}}(r,i)$ of our result and of the composite spectrum \citep{van01} are generally consistent.
\cite{van01} presented the geometric and arithmetic composite spectra, but their difference is very small (their spectral index is $\alpha_{\nu}=$ $-0.44$ and $-0.46$, respectively, with the uncertainty $\sim 0.1$ due to the spectrophotometric calibration) and they have almost the same $b^{\text{ratio}}(r,i)-$redshift relation.
Power-law index $\alpha_{\nu}=-0.44$ (shown as dotted line in upper panel of Figure~\ref{fig:comparison}) is consistent with $b^{\text{ratio}}(r,i)$, and as we have shown in Figure~\ref{fig:b_ugriz}, all $b^{\text{ratio}}(s,l)$  are also consistent with $\alpha_{\nu}=-0.44$.

On the other hand, $b(r,i)$ of the regression results and the geometric composite difference spectrum \citep{wil05} (middle panel of Figure~\ref{fig:comparison}) seem to be inconsistent in that $b(r,i)$ of the regression results are generally lower than that of the composite difference spectrum for any redshift.
This indicates that their composite geometric difference spectrum is steeper than our sample mean color of the difference spectra.
We note that their difference spectra composed by taking geometric mean and arithmetic mean are largely different.
The authors did not specify the reasons for this large discrepancy.
They rely on the geometric composite spectrum when inferring the continuum spectral index $\alpha_{\nu}^{\text{dif}}=0.0$, and we should compare our result with the geometric one when inferring the continuum variability.
Also, we note that their spectra are not corrected for Galactic extinction  \citep[see][]{aba04,wil05}.
The composite spectra with no correction for Galactic extinction obtained by \cite{wil05} has the spectral index $\alpha _{\lambda }^{\text{com}} = -1.35$, which is steeper (redder) than the dereddened composite spectrum by  \cite{van01} ($\alpha _{\lambda }^{\text{com}} = -1.56$).
So, the composite flux difference spectrum with $\alpha _{\lambda}^{\text{dif}} = -2.0$ with no correction for Galactic extinction obtained by  \citep{wil05} means
that the intrinsic composite flux difference spectrum should have flatter (bluer) spectral index (i.e., $\alpha _{\lambda }^{\text{dif}}$ $<$ $-2.0$), which is consistent with our result $\alpha _{\lambda}^{\text{dif}}$ $\sim$ $-7/3$ or $\alpha _{\nu}^{\text{dif}}$ $\sim$ $+1/3$.
In the bottom panel of Figure~\ref{fig:comparison} we show $b(r,i)$ with no correction for Galactic extinction as a function of redshift.
In this plot, although bumpy features in the $b(r,i)-$redshift relation (which correspond to the emission lines variability and are discussed in Sections~\ref{sec:5} and \ref{sec:6}) of the regression results are still not in good agreement with that of the geometric composite difference spectrum \citep{wil05}, it is clear that our result and that of the geometric composite difference spectrum \citep{wil05} are approximately consistent in that they have similar average power-law value.
Thus, a large fraction of the discrepancy between our result and that of \cite{wil05} seen in the middle panel of Figure~\ref{fig:comparison} can be attributed to a correction or no-correction for the Galactic extinction.
It also indicates that the composite flux difference spectrum is not so different in spite of the large difference in time scale between ours and that of  \citep[][; see Figure~\ref{fig:ri_timescale}]{wil05}.

Hereafter, we refer the approximate form of power-law continuum flux difference spectrum as
\begin{equation}
f_{\nu}^{\text{dif}} \propto \nu ^{+1/3}
\label{single_powerlaw}
\end{equation}
(i.e., $\alpha_{\nu}^{\text{dif}}\sim +1/3$).
This power-law index is well known as the standard disk model prediction \citep{lyn69,sha73}.
\cite{tom06} and  \cite{lir11} concluded that the flux difference spectra in near infrared (NIR) wavelengths for their AGN samples are well represented by a combination of the dusty torus component (Blackbody spectrum) and the accretion disk component ($\alpha_{\nu}^{\text{dif}}\sim +1/3$) on the assumption that the flux difference spectrum conserves the spectral shape (spectral index) of the underlying accretion disk spectrum \citep[see also][]{pal96,col99}.
Note that the observations of the polarized flux spectra of AGNs $-$ which is assumed to be produced by electron scattering interior to the dust sublimation radius and is regarded as a copy of the spectrum originating in the accretion disk $-$ also revealed the underlying continuum spectral shape consistent with $\alpha_{\nu}=+1/3$ \citep[$\alpha_{\nu}=+0.44\pm0.11$, ][]{kis08}.
Our result can be considered to be the UV$-$optical extension of the previous results on the flux difference spectrum obtained in the optical$-$NIR wavelengths \citep{tom06,lir11}, because the dust in the nuclear region cannot become higher temperature than the dust sublimation temperature (${T}_{\text{sub}}\sim 1500$K) and the accretion disk component dominate in UV$-$optical wavelengths.

In Section~\ref{sec:4_3}, we discuss the validity of the accretion disk model in UV$-$optical wavelengths.

\subsection{Comparison with Standard Accretion Disk Model}
\label{sec:4_3}

 \cite{per06} fitted the composite flux difference spectrum presented by  \cite{wil05} with a standard thermal accretion disk model  \citep{sha73}, in which the luminosity varies following the changes in accretion rate from one epoch to the next, and concluded that most of the UV$-$optical variability observed in quasars may be due to processes involving changes in mass accretion rates.
\cite{sak11} investigated this model for 10 SDSS quasars individually, and also concluded that the model can explain the UV-optical variability quantitatively.
Definitions of the model adopted by  \cite{per06} and \cite{sak11} are slightly different in that the former depends on the first order Taylor expansion about the average characteristic temperature ${T}^*$ \citep[see][]{per06}, but the predicted flux difference spectra are identical for the modest variation amplitude, which is valid for most of quasars.

\begin{figure}[tbp]
 \begin{center}
 \hspace{-0.7cm} \vspace{-0.1cm}
 \includegraphics[clip, width=3.15in]{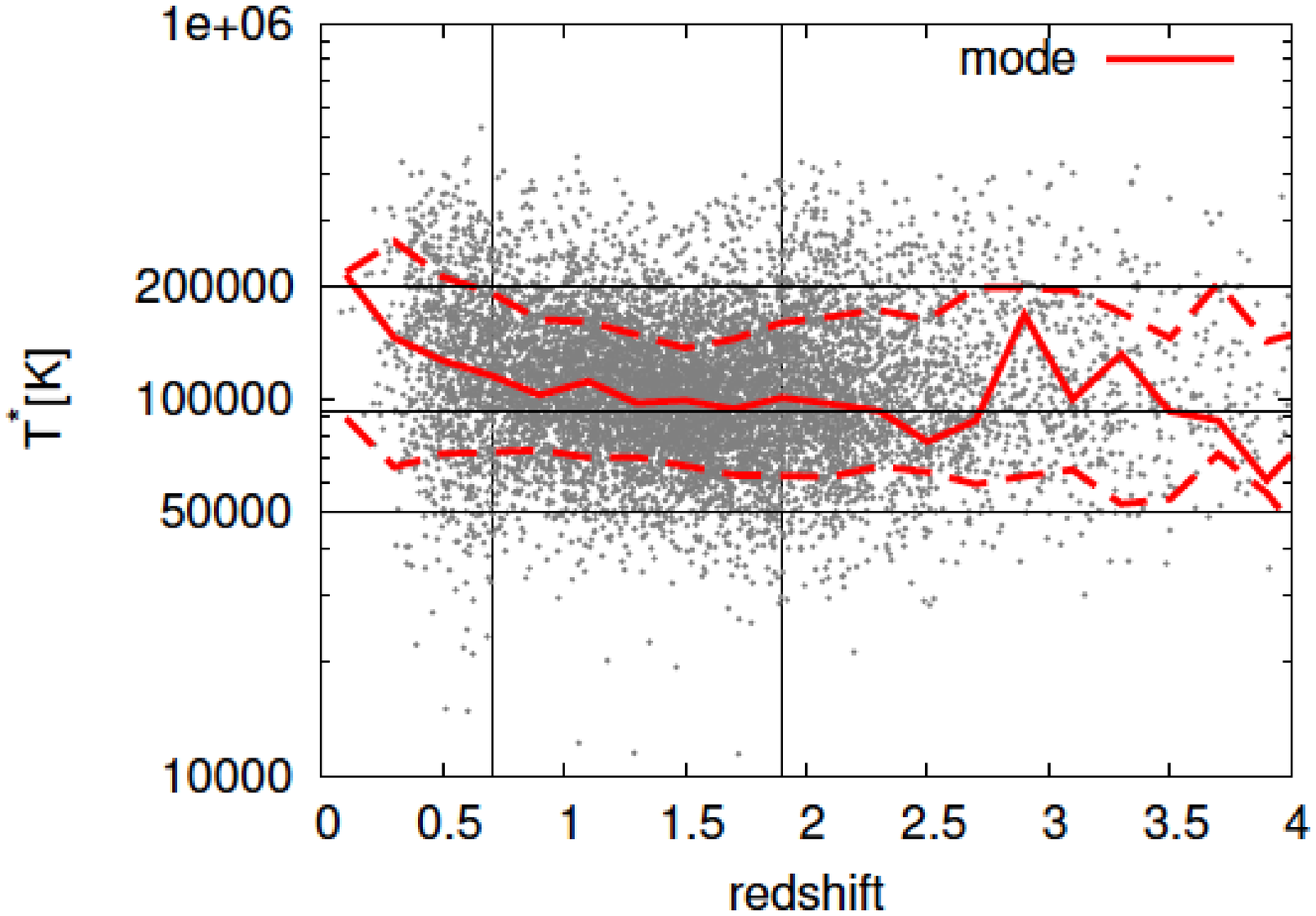}
 \vspace{-0.3cm}
 \includegraphics[clip, width=3.2in]{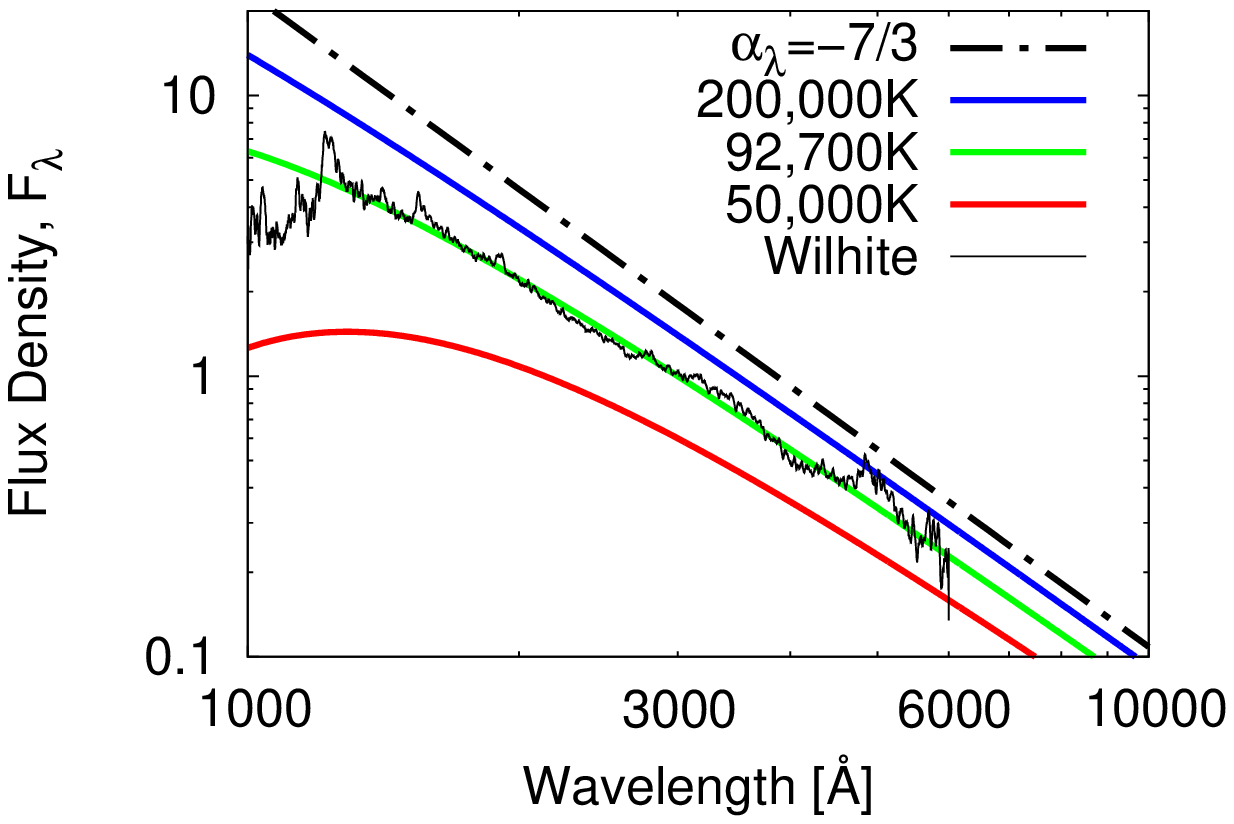}
 \end{center}
 \caption{Upper panel: characteristic temperature of SDSS Stripe~82 quasars calculated by Equation~(\ref{characteristic_temp}) with the use of a quasar property catalog \citep{she11}.
Vertical thin solid lines indicate redshift ranges in which bolometric luminosities are computed in different ways: from
 $L_{5100}$ ($z$ $<$ 0.7), $L_{3000}$ (0.7 $<$ $z$ $<$ 1.9), $L_{1350}$ ($z$ $>$ 1.9) using bolometric corrections $BC_{5100} = 9.26$, $BC_{3000} = 5.15$, and $BC_{1350} = 3.81$, respectively \citep{she11}.
A horizontal thin solid line indicates ${T}^*=$ 200,000~K, 92,700~K   \citep[best-fit value obtained by][]{per06}, 50,000~K.
The solid curve with dashed outer quartiles (colored red in the electronic edition) indicate the mode and the 1$\sigma$ range for each redshift bin ($\Delta z$ $=$ 0.2).
Bottom panel: model flux difference spectra $f_{\nu}^{\text{dif}}$  \citep{per06}, calculated for ${T}^* = 200,000$~K (bluest), $92,700$~K, and $50,000$~K (reddest).
A dash-dotted line is a power-law with $\alpha_{\lambda}=-7/3$.
For comparison, the geometric composite flux difference spectrum by \cite{wil05} is shown as a thin line.
Spectra are arbitrarily scaled.}
\label{fig:characteristic_temperature}
\end{figure}

In the formalization of \cite{per06}, the model flux difference spectrum has two free parameters: a normalization constant and the average characteristic temperature ${T}^*$.
One of the advantages of focusing on the color$-$redshift relation is that it does not depend on the inclination angle \citep[including limb-darkening effect, e.g.,][]{phi86}.
When inferring the color of the flux difference spectrum (observable $b(s,l)$ in the present work), this model requires only one parameter ${T}^{*}$ because the normalization constant does not affect the color of the difference spectrum.
${T}^*$ is defined as
\begin{eqnarray}
T^*&=&\left(\frac{3GM_{\text{BH}}\dot{M}}{8\pi \sigma _S (3R_S)^{3} }\right)^{1/4}\\
&=&90,600 \text{K} \left(\frac{\epsilon}{1/12}\right)^{-1/4}\left( \frac{M_{\text{BH}}}{10^{9}M_{\odot}} \right)^{-1/4}\left( \frac{\text{ER}}{0.1} \right)^{1/4}
\label{characteristic_temp}
\end{eqnarray}
where $M_{\text{BH}}$ is the black hole mass, $\dot{M}$ is the mass accretion rate, $\sigma_{S}$ is the Stefan-Boltzmann constant, and K denotes the kelvin.
The disk inner edge is assumed to be at the inner stable circular orbit ($R_{\text{in}}=3R_S$, where $R_S$ is Schwarzschild radius).
Eddington luminosity is defined as $L_{\text{Edd}}=4\pi G M_{\text{BH}}m_p c/\sigma_{T}$ (where $m_p$ is the proton mass and $\sigma_{T}$ is the Thomson cross section) and Eddington ratio, denoted ``ER'', is given as ER $\equiv L_{\text{bol}}/L_{\text{Edd}}$.
Radiative efficiency $\epsilon$, defined as $L_{\text{bol}}=\epsilon \dot{M}c^2$, is fixed\footnote{In the (non-relativistic) standard disk model, the radially integrated disk luminosity is $L_{\text{bol}}=(GM_{\text{BH}}\dot{M})/(2R_{\text{in}})=(1/12)\dot{M}c^2$, then $\epsilon=1/12$  \citep{kat08}.} to $\epsilon=1/12$.
Changes in $\dot{M}$ lead to the changes in ${T}^*$ as ${T}^*\propto \dot{M}^{1/4}$ and then to the changes in flux.
\cite{per06} fitted their continuum variability model to the geometric mean flux difference spectrum of  \cite{wil05}, and obtained ${T}^*=92,700$~K.
The upper panel of Figure~\ref{fig:characteristic_temperature} shows the characteristic temperature of our quasar samples calculated by Equation~(\ref{characteristic_temp}) using the fiducial virial black hole mass and Eddington ratio listed in quasar property catalog \citep{she11}.
${T}^*=92,700$~K, obtained by \cite{per06}, is very reasonable for SDSS quasars.

\begin{figure}[!t]
\centerline{
\includegraphics[clip, width=2.7in]{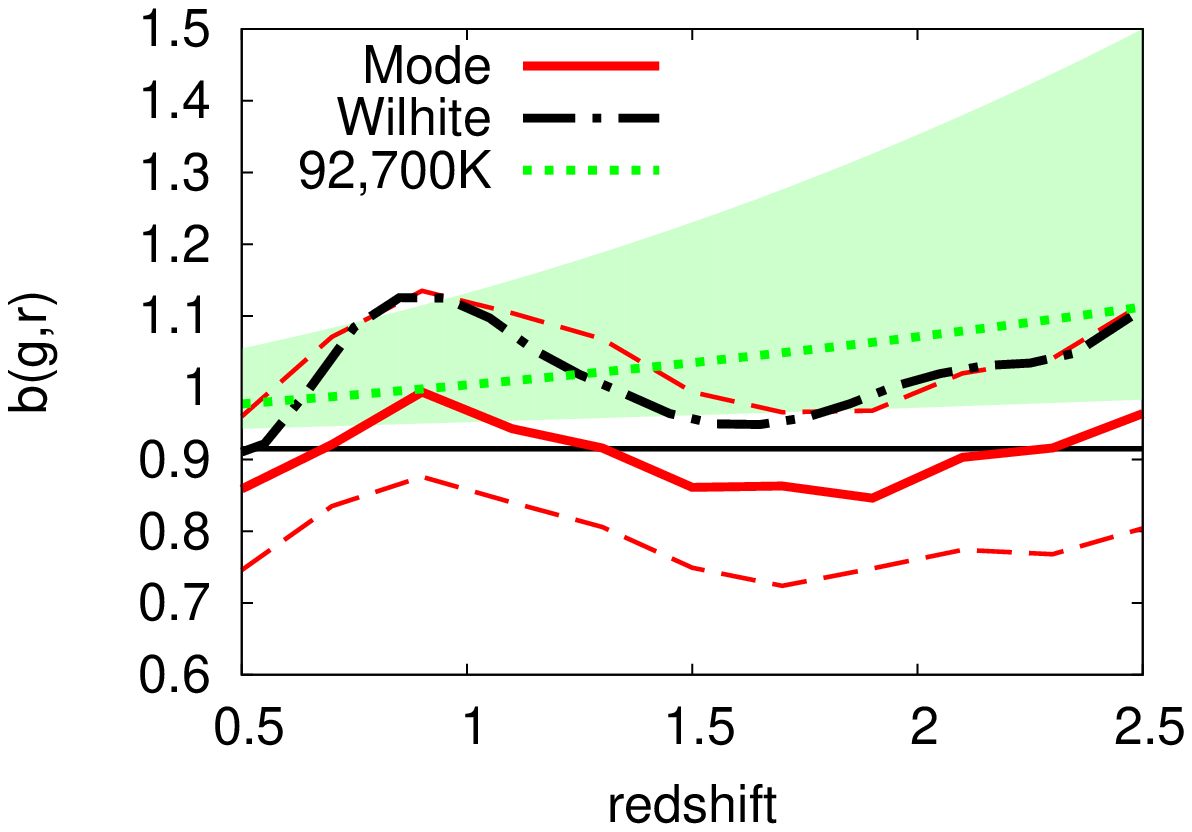}
}
\centerline{
\includegraphics[clip, width=2.7in]{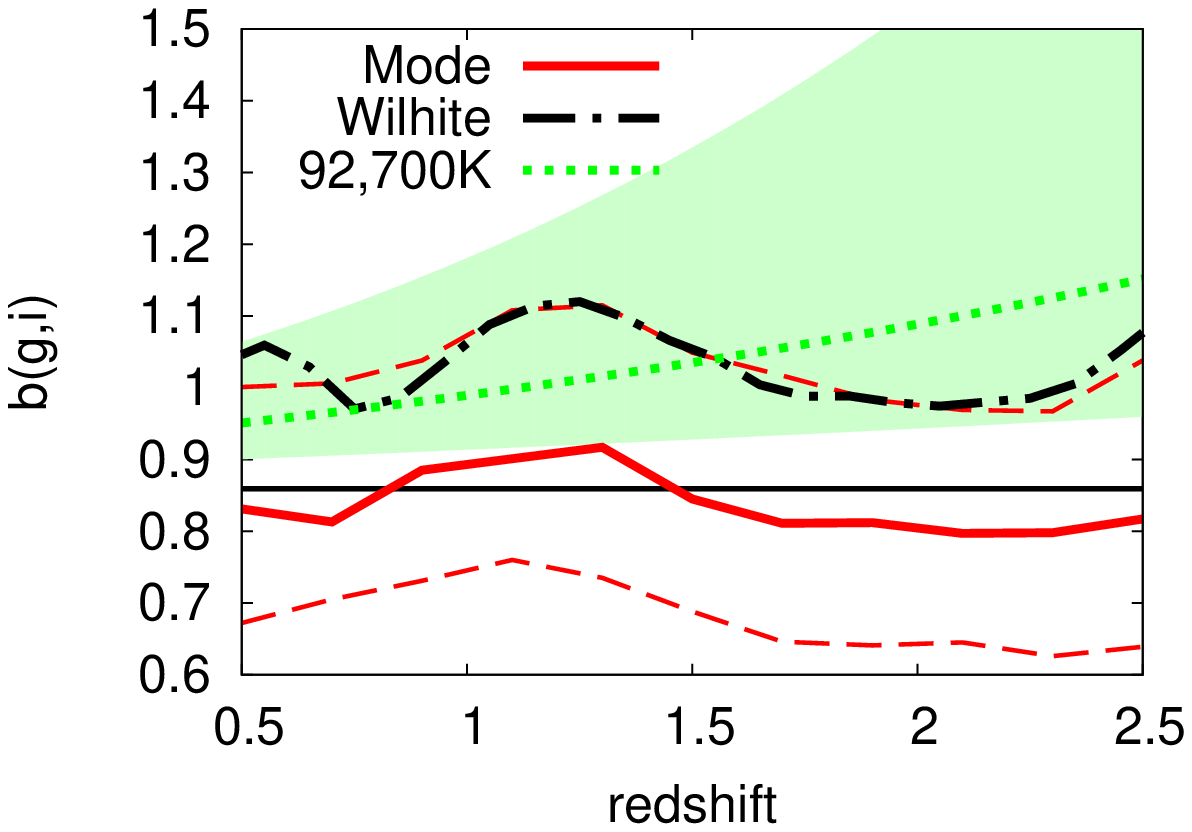}
}
\centerline{
\includegraphics[clip, width=2.7in]{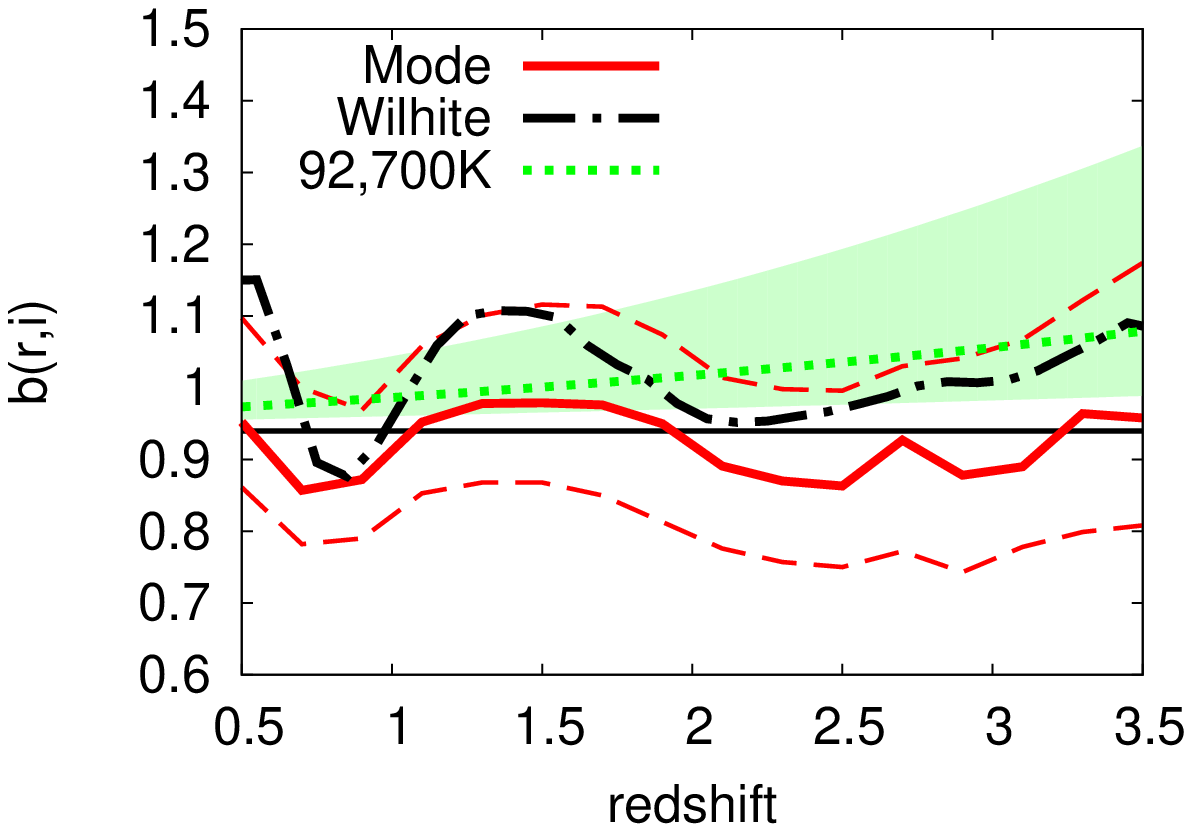}
}
\caption{Comparison of the observed $b(s,l)$ (corrected for Galactic extinction) with prediction by a standard thermal accretion disk model \citep{per06} in $b(s,l)$-redshift space.
Dotted curves (colored green in the electronic edition) indicate a standard thermal accretion disk model with changing mass accretion rate, assuming that the average characteristic disk temperature is ${T}^*=92,700$K  \citep{per06}.
The filled region (colored light green in the electronic edition) corresponds to the disk temperature ranges from $T^*=$50,000~K to 200,000~K (see Figure~\ref{fig:characteristic_temperature}).
The solid curves with dashed outer quantile (colored red in the electronic edition) are the same as those in Figure~\ref{fig:b_ugriz}.
The dash-dotted curves are calculated from the geometric composite flux difference spectrum (the same as those in Figure~\ref{fig:comparison}).
Horizontal thin solid lines are the color of the power-law difference spectrum with $\alpha_{\nu}^{\text{dif}}=+1/3$.}
\label{fig:comparison_pereyra}
\end{figure}

This model predicts $\alpha_{\nu}^{\text{dif}}=1/3$ around optical to near infrared wavelengths \citep[e.g.,][]{sak10}.
This spectral index is well known as the long wavelength limit for a standard thermal accretion disk model \citep{sha73}.
At the long wavelength limit, the accretion disk model spectrum with varying accretion rate turns out to be simple scaling, thus the flux difference spectrum predicted in this model also has $\alpha_{\nu}^{\text{dif}}\sim +1/3$.
Because we obtained $\alpha_{\nu}^{\text{dif}}\sim +1/3$ as shown in Figure~\ref{fig:b_ugriz}, this model is approximately valid.
However, we should be careful because:
\begin{itemize}
\item The long wavelength limit approximation is not valid for the wavelength range of SDSS photometry for high-redshift quasars.
In UV wavelengths, $f_{\nu}^{\text{dif}}$ predicted by this model becomes gradually steeper (redder) because the UV spectral range corresponds to the turnover spectral region of the blackbody radiation from the accretion disk inner boundary at the inner stable circular orbit (i.e., the hottest part of the disk), as seen in the lower panel of Figure  \ref{fig:characteristic_temperature}.
\item  \cite{per06} used the composite difference spectrum not corrected for Galactic extinction  \citep{wil05}, and the effect of the non-correction is not investigated in their paper.
The same is true for \cite{sak11}, in which the correction for the Galactic extinction was not properly applied.\footnote{\cite{sak11} mentioned that they used the point-spread function magnitude corrected for Galactic extinction obtained from the SDSS database, but we found that the light curves shown in their figures are not properly corrected for it.}
It is possible that the non-correction for Galactic extinction leads to the wrong conclusion.
\end{itemize}

So, it is worth comparing our (Galactic-extinction corrected) result with the model prediction to clarify the validity of the standard disk model.
We compare our results with the model proposed by  \cite{per06} in $b(r,i)-$redshift space as shown in Figure~\ref{fig:comparison_pereyra}.
In Figure~\ref{fig:comparison_pereyra}, Pereyra's model (with their best-fit value of $T^*=$92,700K) is actually in agreement with the Wilhite's composite flux difference spectrum.
However, the model prediction is apparently steeper ($b(r,i)$ is larger) than our result.
This discrepancy is mainly due to whether or not Galactic extinction is corrected (see middle and bottom panel in Figure~\ref{fig:comparison}).
In Figure~\ref{fig:comparison_pereyra}, we also plot the model prediction range in $b(s,l)$-redshift space, which is from $T^*=50,000$~K (upper boundary) to $200,000$~K (lower boundary).
The model prediction is apparently not consistent with the $b(s,l)-$redshift region obtained in our analyses, in that the model predicts larger $b(s,l)$ values and larger spread in $b(s,l)-$redshift space than observed values particularly in the high redshift range.

These indicate that the intrinsic composite flux difference spectrum of quasars is actually flatter (bluer) than the prediction of the accretion disk model, particularly in the UV wavelength range.
In other words, we cannot see the model-predicted ``UV turnover'' feature \citep[e.g.,][]{sha05} in the flux difference spectrum.
This result is qualitatively consistent with \cite{sch12}, but our method enables us to examine the difference between observed variability and the model prediction quantitatively.

We are able to make the model flux difference spectrum flatter (up to $f_{\nu}^{\text{dif}}\propto \nu^{1/3}$ at all UV$-$optical wavelengths) if we assume the disk characteristic temperature to be , for example, ${T}^* > 200,000$~K, but this is not valid for most of our quasar samples as shown in the upper panel of Figure~\ref{fig:characteristic_temperature}.
Beyond the standard accretion disk model \citep{sha73}, several more sophisticated accretion disk models are proposed \citep[see, e.g.,][]{hub00,dav07}.
However, accretion disk model spectra based on non-LTE atmosphere calculations \citep{dav07} usually have redder color than the standard thermal accretion disk model because of opacity effects  \citep[e.g.,][]{hub00} and this leads to the redder color of the flux difference spectrum as shown in \cite{sch12}.
Although radiative transfer calculations for a disk$+$corona configuration (rather than those for a bare disk above) result in a weaker ``UV turnover'' \citep{kaw01}, the expected spectra are redder than the long wavelength limit of the standard disk as well.
In addition, the accretion disk of quasars or luminous AGNs is thought to be seen almost face-on, so the relativistic effects for the color of the observed spectrum is probably negligible  \citep{hub00}.
In short, it is difficult to explain the flux difference spectrum with $\alpha_{\nu}^{\text{dif}} \sim 1/3$ by any of the existing accretion disk models.

However, it is true that the reasonable range of physical parameters related to the accretion disk model ($M_{\text{BH}}$ and Eddington ratio) can qualitatively explain the absolute continuum flux and flux variation amplitude as indicated in  \cite{sak11} and \cite{gu13}, and $\alpha_{\nu}^{\text{dif}}$ obtained here can be naturally attributed to the well-known long wavelength limit value predicted in standard accretion disk models.
Thus, we can only conclude that the continuum spectral variability in quasars cannot be explained by the accretion disk model with varying mass accretion rate, but it is probably related to thermal accretion disk itself.
Finally, we note that it seems to be difficult to explain the large coherent flux variation within UV$-$optical wavelength range (e.g., a strong linear flux$-$flux correlation in each individual quasar) by the accretion disk local fluctuations or hot spot scenarios \citep[e.g.,][]{dex11,sch12,meu13}.

\subsection{Brief Summary of This Section}
\label{sec:4_4}

In summary, our conclusions in this section are:
\begin{enumerate}
\item We show that the continuum component of the flux difference spectra is (in average) well approximated by a power-law shape with $\alpha_{\nu}^{\text{dif}}\sim +1/3$, and it is bluer than the spectroscopic composite flux difference spectrum  ($\alpha_{\nu}^{\text{dif}}=0.0$) obtained by \cite{wil05}.
We conclude that the discrepancy is due to the Galactic extinction; the light curves used in our analyses are corrected for Galactic extinction but the spectra used in \cite{wil05} are not.
If we derive the color of the flux difference spectrum without the correction, our result shows consistent color and rest-frame wavelength dependence of the flux difference spectrum with \cite{wil05}, in spite of the large difference in time scale (Figure~\ref{fig:ri_timescale}).
\item We compare the $b(s,l)-$redshift relation with ``the standard disk with varying mass accretion rate model'' \citep{per06}.
One of the advantages of our method is that it is based on the direct comparison of observable color of the difference spectrum ($b(s,l)$) with the model predicted difference spectrum, and does not require additional models or assumptions for non-variable spectral components.
We concluded that the flux difference spectrum is flatter (bluer) than the model prediction.
In particular, we confirm that the model-predicted ``UV turnover'' is not seen in the flux difference spectrum.
This is surprising because not only the standard accretion disk model \citep{sha73} but also any accretion disk models cannot produce such a flat spectrum as $\alpha_{\nu}^{\text{dif}}\sim +1/3$.
This result is qualitatively consistent with \cite{sch12}, but our method enables us to examine the difference between observed variability and the model prediction quantitatively.
\end{enumerate}

\section{Emission line variability: The origin of the redshift dependence of the color of the flux difference spectrum}
\label{sec:5}

\begin{figure}[tbp]
 \begin{center}
  \includegraphics[clip, width=3.4in]{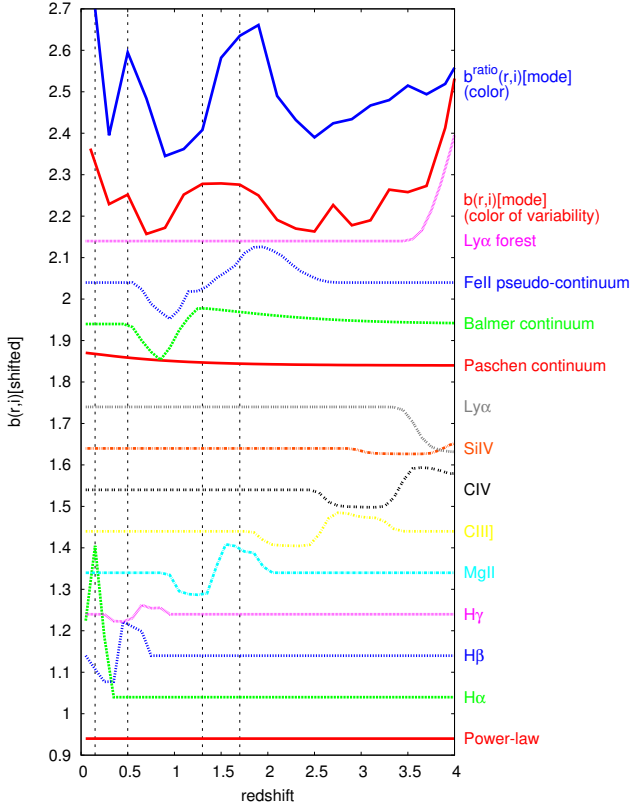}
 \end{center}
 \caption{$b(r,i)$ as a function of redshift.
Each curve (except for the top two curves) is calculated by Equation~(\ref{b_cal}) assuming a difference spectrum $f^{\text{dif}}_{\nu}$ composed of power-law continuum and one additional spectral component: the broad emission lines, Balmer and Paschen continuum, or Ly$\alpha $ forests (Equations~(\ref{model_dif_spec}) and (\ref{model_dif_spec_cont})).
The solid curve labeled ``$b^{\text{ratio}}(r,i)$[mode]'' (colored blue in the electronic edition) is the modal color of the time-averaged spectrum $b^{\text{ratio}}(r,i)$ as a function of redshift.
The solid curve labeled ``$b(r,i)$[mode]'' (colored red in the electronic edition) is the modal color of the difference spectrum ($b(r,i)$) as a function of redshift.
Curves (other than the ``power-law'' model) are shifted vertically, and the offset is taken to be 0.1 from adjacent curves.}
 \label{fig:itiran}
\end{figure}

\begin{figure}[tbp]
 \begin{center}
  \includegraphics[clip, width=3.4in]{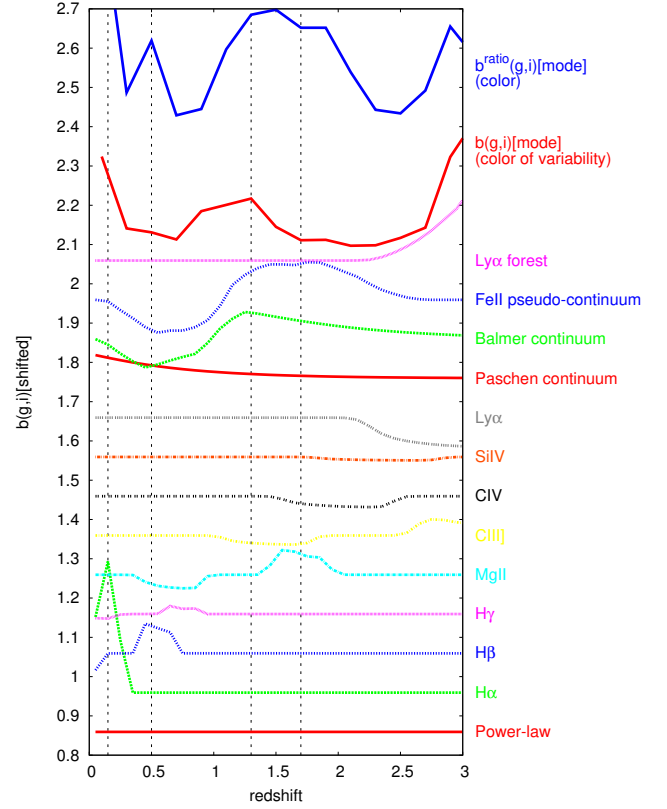}
 \end{center}
 \caption{Same as Figure~\ref{fig:itiran}, but for $b(g,i)$ and $b^{\text{ratio}}(g,i)$.}
 \label{fig:itiran_2}
\end{figure}

The flux difference spectra are composed of the continuum component (accretion disk emission) and the BELs component.
The regression slope $b(s,l)$ can be related to the difference spectrum as in Equation~(\ref{b_cal}), and as we have seen in Section~\ref{sec:3_3} (Figure~\ref{line_contami_ri}), emission line variability on the power-law continuum variability makes bumpy features in $b(s,l)-$redshift space.
Many of the features in the color-redshift (i.e., $b^{\text{ratio}}(s,l)-$redshift, or $b(s,l)-$redshift) relations are caused by more than one feature in the quasar spectrum \citep{ric01}.
We are able to infer the contribution of each of the emission line variability to the flux difference spectrum by examining the $b(s,l)-$redshift relation.

We show the effects of the variability of each of the prominent BEL on $b(r,i)$ and $b(g,i)$  as a function of redshift in Figures~\ref{fig:itiran} and \ref{fig:itiran_2}, respectively.
``Emission lines'' contain BELs, Balmer and Paschen continuum emission (BaC and PaC), and \ion{Fe}{2}  pseudo-continuum emission.
They are calculated assuming
\begin{equation}
f^{\text{dif}}_{\lambda}=f^{\text{dif}}_{\lambda}(\text{continuum})+0.05\times f^{\text{dif}}_{\lambda}(\text{line})
\label{model_dif_spec}
\end{equation}
where $f^{\text{dif}}_{\lambda}(\text{continuum})$ is the power-law continuum variable component (Equation~(\ref{single_powerlaw})) normalized to 1 at 3000\AA \ , and given as 
\begin{equation}
f^{\text{dif}}_{\lambda}(\text{continuum}) = \left(\frac{\lambda}{3000\text{\AA}}\right)^{-7/3}
\label{model_dif_spec_cont}
\end{equation}
$f^{\text{dif}}_{\lambda}(\text{line})$ includes $f^{\text{dif}}_{\lambda}(\text{BEL})$,
$f^{\text{dif}}_{\lambda}(\text{BaC})$, $f^{\text{dif}}_{\lambda}(\text{PaC})$, and $f^{\text{dif}}_{\lambda}(\text{\ion{Fe}{2}})$.
The factor ``$0.05$'' in Equation~(\ref{model_dif_spec}) is an overall normalization constant for emission line components to a power-law continuum component and is determined to approximately reproduce the amplitude of bumpy features in the observed $b(s,l)-$redshift relation.
The analytical expressions for calculating these components, and the IGM attenuation model (labeled ``Ly$\alpha$ forest'') are summarized in Appendix A.
In Figures~\ref{fig:itiran} and \ref{fig:itiran_2}, the relative strengths of these components are taken into consideration as described in Appendix A.

Here, we discuss several features of $b^{\text{ratio}}(r,i)-$redshift relation and $b(r,i)-$redshift relation qualitatively.
We focus on the case of $(r,i)$ band pair, but the color$-$redshift relation of other band pairs can be interpreted in the same way.

By comparing $b(r,i)-$redshift relation with $b^{\text{ratio}}(r,i)-$redshift relation in Figure~\ref{fig:itiran}, we are able to identify several similarities and differences in the features.
Features of Balmer emission lines and Balmer continuum are seen in both $b(r,i)$ and $b^{\text{ratio}}(r,i)$.
The most significant difference between $b(r,i)$ and $b^{\text{ratio}}(r,i)$ exists in features of the \ion{Fe}{2} pseudo-continuum and the \ion{Mg}{2} emission line.
As was done in \cite{ric01}, we can identify each of the significant features in the $b^{\text{ratio}}(r,i)$-redshift relation and $b(r,i)-$redshift relation for each redshift range as below:
\begin{itemize}
\item $z$ $\sim$ 0.1 to 0.2 --- H$\alpha$ is in $i$-band and makes $b^{\text{ratio}}(r,i)$ and $b(r,i)$ redder than the average power-law value.
Also, the Paschen continuum emission possibly makes the color slightly redder \citep[e.g.,][]{lan11}.
\item $z$ $\sim$ 0.3 --- The presence of H$\beta$ in $r$-band drives $b^{\text{ratio}}(r,i)$ and $b(r,i)$ blueward.
\item $z$ $\sim$ 0.5 --- $b^{\text{ratio}}(r,i)$ and $b(r,i)$ moves back to red color while H$\beta$ is in $i$-band.
\item $z$ $\sim$ 0.7 --- Balmer continuum enters the $r$-band making $b^{\text{ratio}}(r,i)$ and $b(r,i)$ bluer.
\item $z$ $\sim$ 0.9 --- \ion{Fe}{2} and \ion{Mg}{2} fill the $r$-band making $b^{\text{ratio}}(r,i)$ bluer.
However, these features seem to be absent in $b(r,i)$.
\item $z$ $\sim$ 1.4 to 1.5 --- $b^{\text{ratio}}(r,i)$ makes a sharp transition from blue to red as \ion{Mg}{2} leaves $r$-band and enters $i$-band.
This transition seems to be absent in $b(r,i)$
\item $z$ $\sim$ 1.8 --- A small hump is caused as both \ion{Mg}{2} and \ion{Fe}{2}  push $b^{\text{ratio}}(r,i)$ redward.
Again, this feature is absent in $b(r,i)$.
\item $z$ $\sim$ 2.0 to 2.5 --- $b^{\text{ratio}}(r,i)$ is driven back to the blue as \ion{Fe}{2} leaves $i$-band.
$b(r,i)$ also becomes blue, probably because of the contribution from \ion{C}{3}] in $b(r,i)$.
Note that there exist many kinds of weaker high-ionization lines between \ion{C}{3}] and \ion{C}{4} that we do not show (e.g.,\ion{Fe}{3}, \ion{Al}{3}, \ion{N}{3}], and \ion{N}{4}), and they also make $b^{\text{ratio}}(r,i)$ and $b(r,i)$ bluer.
\item $z$ $\sim$ 2.6 to 3.4 --- \ion{C}{4} and  \ion{C}{3}] offset each other when the former is in $r$-band and the latter is in $i$-band, and keeps $b^{\text{ratio}}(r,i)$ and $b(r,i)$ blue.
\item $z$ $>$ 3.4 --- Ly$\alpha$ and \ion{C}{4} offset each other during the period when the former is in $r$-band and the latter is in $i$-band.
At higher redshift, $b^{\text{ratio}}(r,i)$ and $b(r,i)$ rises rapidly as the Ly$\alpha$ forest or Lyman-limit systems absorb $r$-band flux (see Figure~\ref{fig:igm_transmissivity}).
\end{itemize}

\begin{figure}[tbp]
 \begin{center}
  \includegraphics[clip, width=3.4in]{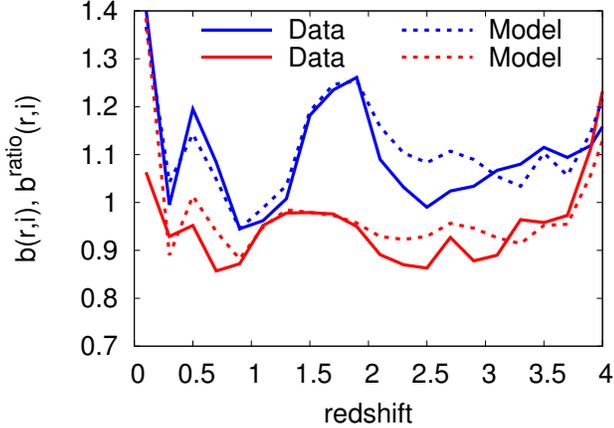}
  \vspace{-0.5cm}
 \end{center}
 \caption{Comparison of observed $b^{\text{ratio}}(r,i)$ (upper solid line, colored blue in the electronic edition) and $b(r,i)$ (lower solid line, colored red in the electronic edition) with those of model spectra (dotted lines) as a function of redshift .
 The model spectra are the sum of a power-law component ($\alpha_{\lambda}=-1.56$ and $\alpha^{\text{dif}}_{\lambda}=-7/3$ for $b^{\text{ratio}}(r,i)$ and $b(r,i)$, respectively) (scaled to 1 at 3000\AA, as Equation~(\ref{model_dif_spec_cont})), and several additional components as below: In the model spectrum for $b^{\text{ratio}}(r,i)$ (upper dotted line, colored blue in the electronic edition), all the emission line components shown in Figure~\ref{fig:itiran}, Balmer and Paschen continuum emission, and \ion{Fe}{2} pseudo-continuum emission are included and the IGM attenuation is applied (the relative strengths of each of the components are the same as Appendix A).
 On the other hand, in the model spectrum for $b(r,i)$ (lower dotted line, colored red in the electronic edition) we exclude \ion{Mg}{2} emission line and \ion{Fe}{2} pseudo-continuum emission.
 Note that there exist many kinds of weaker high-ionization lines between \ion{C}{3}] and \ion{C}{4} that we do not show in Figure~\ref{fig:itiran}/\ref{fig:itiran_2} (e.g.,\ion{Fe}{3}, \ion{Al}{3}, \ion{N}{3}], and \ion{N}{4}), so the model spectra have redder color than the observed color both in $b^{\text{ratio}}(r,i)$ and $b(r,i)$.}
 \label{fig:summed_model}
\end{figure}

In Figure~\ref{fig:summed_model}, we show the comparison of the observed $b^{\text{ratio}}(r,i)$ (upper solid curve) and $b(r,i)$ (lower solid curve) with those of model spectra(dotted curves) as a function of redshift.
The model spectra are the sum of a power-law component ($\alpha_{\nu}=-1.56$ and $\alpha^{\text{dif}}_{\nu}=-7/3$ for $b^{\text{ratio}}(r,i)$ and $b(r,i)$, respectively (scaled to 1 at 3000\AA, as Equation~(\ref{model_dif_spec_cont})), and several additional components as below: In the model spectrum for $b^{\text{ratio}}(r,i)$ (upper dotted curve), all the emission line components shown in Figure~\ref{fig:itiran}, Balmer and Paschen continuum emission, and \ion{Fe}{2} pseudo-continuum emission are included, and the IGM attenuation is applied (the relative strengths of each of the components are the same as Appendix A).
On the other hand, in the model spectrum for $b(r,i)$ (lower dotted curve) we exclude \ion{Mg}{2} emission line and \ion{Fe}{2} pseudo-continuum emission.
Note that there exist many kinds of weaker high-ionization lines between \ion{C}{3}] and \ion{C}{4} that we do not show in Figure~\ref{fig:itiran}/\ref{fig:itiran_2} (e.g.,\ion{Fe}{3}, \ion{Al}{3}, \ion{N}{3}], and \ion{N}{4}), so the model spectra have redder color than the observed color both in $b^{\text{ratio}}(r,i)$ and $b(r,i)$.
Nevertheless, it is clear that the model spectra successfully reproduce the observed bumpy feature in $b(s,l)-$redshift and $b^{\text{ratio}}(s,l)-$redshift relations.
Moreover, it is also clearly shown that the differences between 
$b^{\text{ratio}}(s,l)$ and $b(s,l)$ as a function of redshift exist not only in the continuum power-law spectral index, but also in the relative significance of the emission line components; \ion{Mg}{2} and \ion{Fe}{2} emission line components seem not to play an important role in the $b(s,l)-$redshift relation, although they have significant contribution in the $b^{\text{ratio}}(r,i)-$redshift relation.

In summary:
\begin{enumerate}
\item Utilizing the average $b^{\text{ratio}}(s,l)-$redshift relation, 
we can identify the emission line contamination in the color of the quasars \citep{ric01} as bumpy features in $b^{\text{ratio}}(s,l)-$redshift curves.
Applying this method to the $b(s,l)-$redshift relation, we identify the features due to, for instance, Balmer emission lines and Balmer continuum variability in $b(s,l)-$redshift relation.
\item We show that several features seen in $b^{\text{ratio}}(s,l)-$redshift and $b(s,l)-$redshift relation are similar (e.g., the Balmer series emission), but they look significantly different around the features attributable to the broad \ion{Mg}{2} emission line.
This indicates that the variability of \ion{Mg}{2} emission line and \ion{Fe}{2} pseudo-continuum emission seem to be relatively weaker than other emission line components, such as Balmer series emission lines.
This spectral region (referred to as ``SBB'') also contains the Balmer continuum, \ion{Fe}{2} pseudo-continuum emission, and some other (relatively weak) high-ionization lines, so detailed spectral decomposition analysis is needed to infer the variability of each component, and this is attempted in the next section.
\end{enumerate}

\section{Variability of the Small Blue Bump}
\label{sec:6}

In this section, we focus on the variability of the SBB spectral components.
First we decompose the high resolution composite spectrum of \cite{van01} into spectral components, and compose several model spectra by combining some of the components.
Then, we calculate the $b(r,i)-$redshift relation of each of the model spectrum and compare it with the observed $b(r,i)-$redshift relation.
This procedure enables us to identify the (non-)variable spectral components in the SBB.

\subsection{Spectral Decomposition}
\label{sec:6_1}

\begin{figure}[tbp]
 \begin{center}
  \includegraphics[clip, width=3.0in]{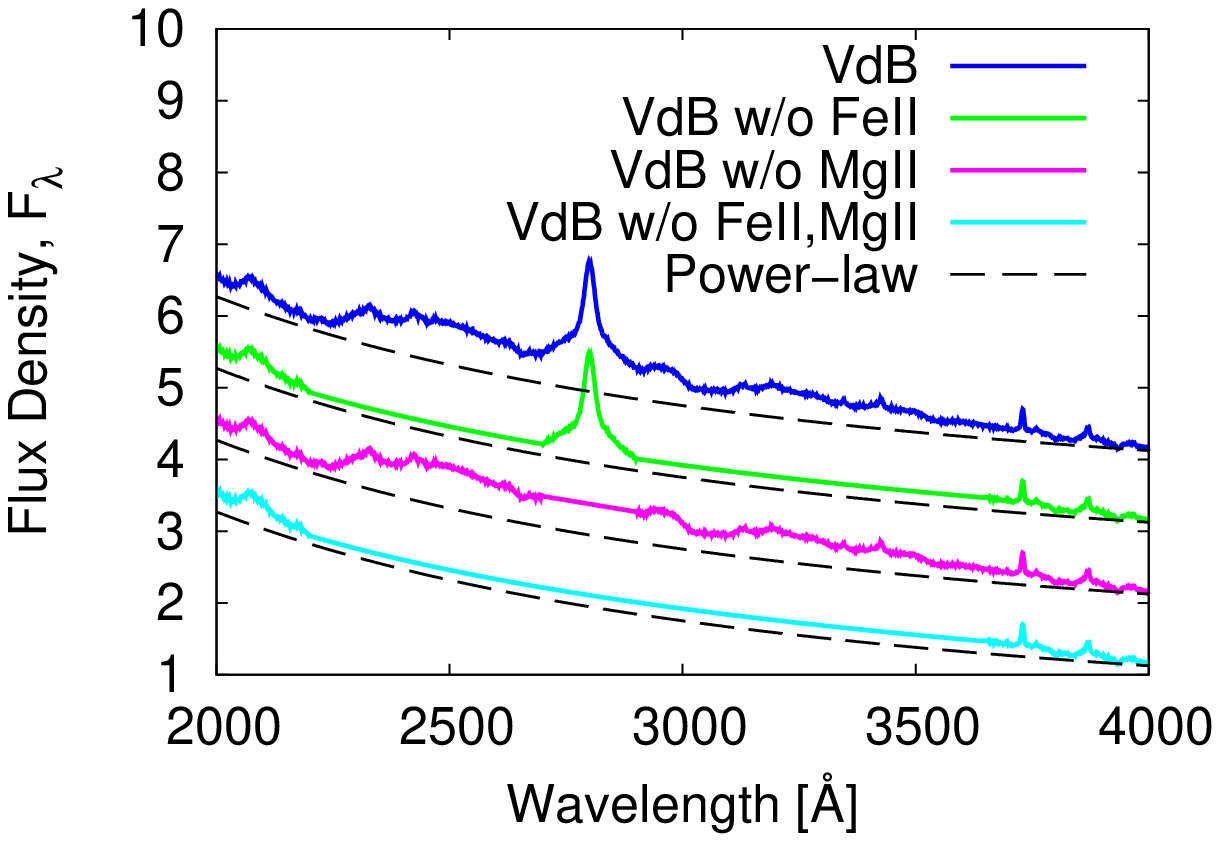}
  \includegraphics[clip, width=3.0in]{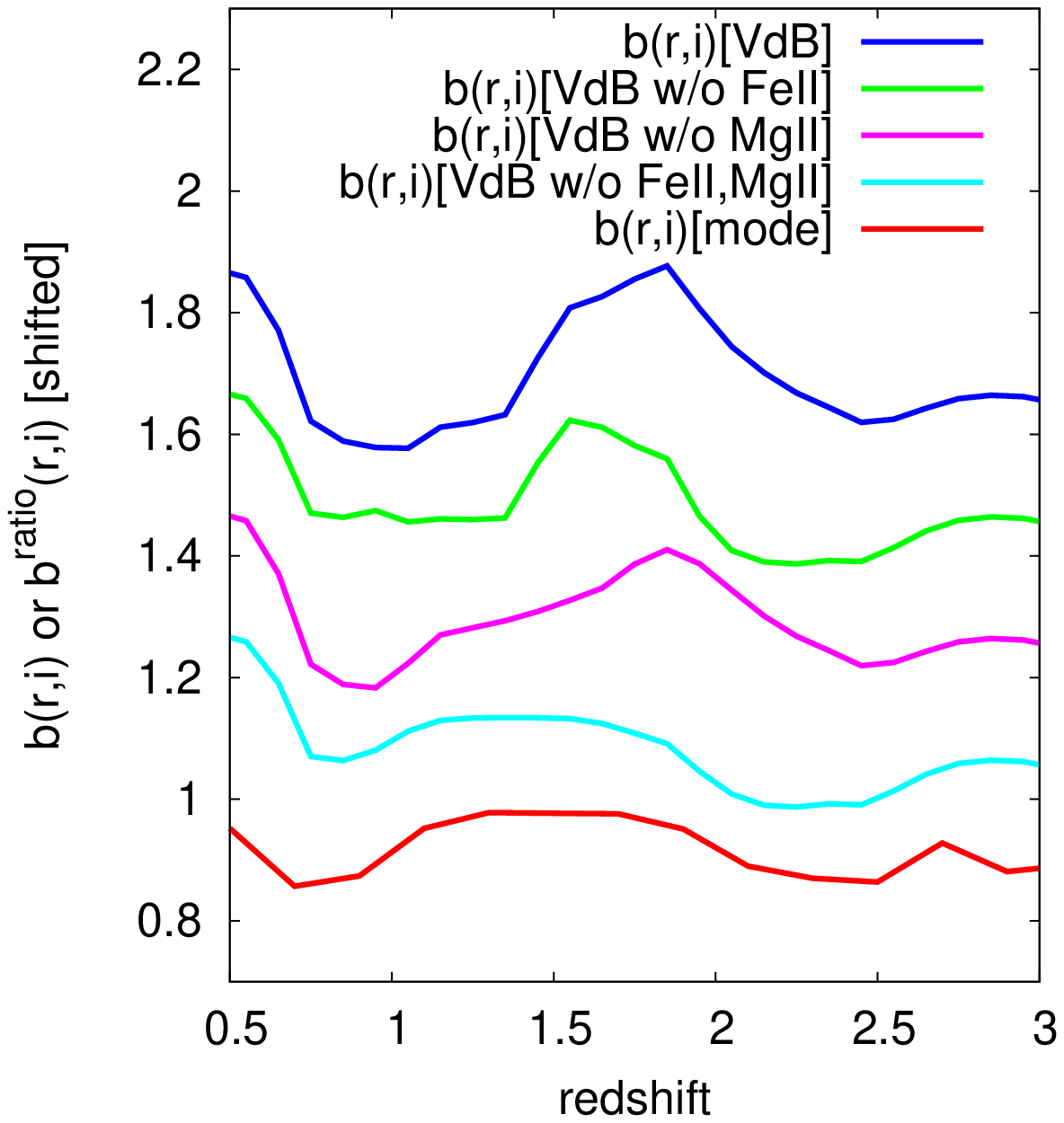}
  \vspace{-0.5cm}
 \end{center}
 \caption{Upper panel: decomposition of the Small Blue Bump (2200\AA \ - 3646\AA) in the composite spectrum  \citep{van01} following  \cite{kur07}.
``VdB'' indicate the Vanden Berk's quasar composite spectrum.
From top to bottom, solid curves indicate the raw composite spectrum (VdB), the \ion{Fe}{2}-subtracted composite spectrum (VdB w/o \ion{Fe}{2}), the \ion{Mg}{2}-subtracted composite spectrum (VdB w/o \ion{Mg}{2}), and the \ion{Fe}{2}-\ion{Mg}{2}-subtracted composite spectrum (VdB w/o \ion{Fe}{2}, \ion{Mg}{2}), respectively (colored blue, green, magenta, and cyan in the electronic edition).
These spectra are shifted by +3.0, 2.0, 1.0, and 0.0 in $y$-axis, respectively.
The continuum power-law index is $\alpha_{\lambda}=-$1.54 and the power-law component is normalized to the flux values in the 4200-4230\AA \ region (dashed lines).
The Balmer continuum is created by Equation~(\ref{bac_cal}) and normalized to the value at 3675\AA.
\ion{Mg}{2} is decomposed by linearly interpolating the spectrum through 2700\AA \ to 2900\AA. 
Bottom panel: the color of the redshifted spectra in upper panel as a function of redshift (upper four curves), compared with the observed modal color of the difference spectrum ($b(r,i)$) as a function of redshift (bottom solid curve, the same as Figure~\ref{fig:b_ugriz}).
From top to bottom, the upper four curves are shifted by +0.6, 0.4, 0.2 and 0.0 in the $y$-axis, respectively (colored blue, green, magenta, and cyan, respectively, in the electronic edition).
The composite spectrum without \ion{Fe}{2} pseudo-continuum and the \ion{Mg}{2} emission line (VdB w/o \ion{Fe}{2},\ion{Mg}{2}), which means this spectrum is composed of a power-law continuum and the Balmer continuum emission, has a similar color$-$redshift relation with the observed $b(r,i)-$redshift curve.
Note that the host galaxy contamination in the VdB composite spectrum is not subtracted.}
 \label{fig:sbb_vandenberk}
\end{figure}

As seen in Figure~\ref{fig:b_ugriz} and discussed in the previous section, the most apparent difference between $b^{\text{ratio}}(s,l)$[color of the time-averaged spectrum] and $b(s,l)$[color of the flux difference spectrum] exists in features of the \ion{Fe}{2} pseudo-continuum and the \ion{Mg}{2} emission line.
The spectral region containing \ion{Fe}{2} pseudo-continuum, Balmer continuum emission, and \ion{Mg}{2} emission line is called SBB \citep{wil85}.
Note that in the actual quasar spectra, higher order Balmer lines are merging to a pseudo-continuum and yielding a smooth rise to the Balmer edge \citep{die03}, which is also contained in SBB.
Here we discuss the SBB variability.

To clarify the difference between $b^{\text{ratio}}(s,l)$ and $b(s,l)$ (as a function of redshift) arising from the complex emission line composition, we decompose the SBB in the composite spectrum \citep{van01} into power-law continuum, \ion{Mg}{2} emission line, \ion{Fe}{2} pseudo-continuum, and Balmer continuum and investigate the color (corresponding to $b^{\text{ratio}}(s,l)$ and $b(s,l)$) as a function of redshift for the composite spectra with and without some SBB components.

 \cite{kur07} decomposed the Vanden Berk's composite spectrum into a power-law continuum, Balmer continuum, and residual.
 They confirmed that the residual was well described by the \ion{Fe}{2}  emission template  \citep{ves01} and a \ion{Mg}{2} emission line.
 We follow their decomposition procedure, which is done in the $2200$-$3646$\AA \ region.
 First, the power-law continuum with $\alpha_{\lambda}=-1.54$ is normalized to the values in the $4200$-$4230$\AA \  region.
 Second, the Balmer continuum is created by Equation~(\ref{bac_cal}) and normalized to the value at 3675\AA \  .
The residual spectrum (the composite spectrum minus a power-law minus Balmer continuum) is composed of \ion{Fe}{2}  pseudo-continuum and a \ion{Mg}{2} emission line.
The \ion{Mg}{2} emission line is decomposed by  simply assuming the \ion{Fe}{2}  pseudo-continuum in 2700\AA \ to 2900\AA \ region is described by linear interpolation of the residual spectrum from 2700\AA \ to 2900\AA.

Using the decomposed components, we define four spectra of the SBB region $f_{\lambda}(\text{SBB})$ as below: 
\begin{itemize}
\item VdB: the composite spectrum of \cite{van01}.
\begin{equation}
f_{\lambda}(\text{SBB})=f_{\lambda}(\text{PL})+f_{\lambda}(\text{BaC})+f_{\lambda}(\text{\ion{Fe}{2}})+f_{\lambda}(\text{\ion{Mg}{2}})
\end{equation}
where ``PL'' denotes a power-law component.
\item VdB w/o \ion{Fe}{2}: VdB minus \ion{Fe}{2}  pseudo-continuum.
\begin{equation}
f_{\lambda}(\text{SBB})=f_{\lambda}(\text{PL})+f_{\lambda}(\text{BaC})+f_{\lambda}(\text{\ion{Mg}{2}}).
\end{equation}
\item VdB w/o \ion{Mg}{2}: VdB minus \ion{Mg}{2} emission line
\begin{equation}
f_{\lambda}(\text{SBB})=f_{\lambda}(\text{PL})+f_{\lambda}(\text{BaC})+f_{\lambda}(\text{\ion{Fe}{2}}).
\end{equation}
\item VdB w/o \ion{Fe}{2} , \ion{Mg}{2}:  VdB minus \ion{Fe}{2}  pseudo-continuum minus \ion{Mg}{2} emission line
\begin{equation}
f_{\lambda}(\text{SBB})=f_{\lambda}(\text{PL})+f_{\lambda}(\text{BaC}).
\end{equation}
\end{itemize}

These spectra are shown in the upper panel of Figure~\ref{fig:sbb_vandenberk}.
Then, we calculate the color of these spectra by Equation~(\ref{b_cal}) or (\ref{b_ratio_cal}) as a function of redshift.
Results are shown in the bottom panel of Figure~\ref{fig:sbb_vandenberk}.
We can see that the Balmer continuum, \ion{Fe}{2} pseudo-continuum, and the \ion{Mg}{2} emission line make distinguishable bump features at different redshift ranges.
The peak in $b(r,i)$[VdB] at $z\sim 1.5$-$1.8$ are made of a \ion{Mg}{2} emission line ($z\sim 1.5$) and \ion{Fe}{2}  pseudo-continuum ($z\sim 1.8$) \citep[in agreement with][]{ric01}, and this feature is not seen in the $b(r,i)$[mode].
The $b(r,i)$[mode] curve resembles the case of ``VdB w/o \ion{Fe}{2}, \ion{Mg}{2}'' (i.e., $f_{\lambda}(\text{SBB})=f_{\lambda}(\text{PL})+f_{\lambda}(\text{BaC})$.
 This indicates that the variability of \ion{Mg}{2} emission line and \ion{Fe}{2} pseudo-continuum are relatively weaker compared to the Balmer continuum emission.

\subsection{\ion{Mg}{2} Emission Line Variability}
\label{sec:6_2}

\ion{Mg}{2} line variability has been examined in several papers \citep[see][and references therein]{woo08,hry13}.
 \cite{goa99a,goa99b} showed that continuum flux and the broad high-ionization emission lines of Ly$\alpha$, \ion{C}{4}, \ion{N}{5}(1240\AA), \ion{He}{2}(1640\AA), and the semi-forbidden lines of the \ion{Si}{3}]$+$\ion{C}{3}] blend varied significantly (a factor of 5 at 1365\AA \ and 3 at 2230\AA \ for continuum, and a factor of 2 for Ly$\alpha$ and \ion{C}{4}), while the low-ionization lines of \ion{Mg}{2} and \ion{Fe}{2} do not vary significantly (less than 7\%) during 11 months of observations for the Seyfert 1 galaxy NGC 3516.
 \cite{goa99a} listed three possible physical reasons explaining an absence of response in the \ion{Mg}{2} emission line: 
(1) the continuum band responsible for the \ion{Mg}{2} emission line \citep[i.e., 600 to 800 eV,][]{kro88} is invariant;
(2) the emission line is insensitive to continuum variations;
and (3) the line-emitting region is physically extended and thus has yet to respond to the observed continuum variations.
(1) was denied in \cite{goa99a} by observational evidence that showed that the amplitude of the continuum variations generally increase toward shorter wavelengths at least up to EUV wavelength range for several AGNs \citep[e.g.,][]{mar97,cag99,hal03}.
Strong soft X-ray variability has been observed in many AGNs, so (1) is definitely not true.
(3) is not true, especially in our sample, with three to four years (in quasar rest-frame) observational duration.
Thus, we are able to expect situation (2) to be responsible for an absence of variability in the \ion{Mg}{2} emission line.
\cite{goa93} defined the line responsivity $\eta $ as a linear coefficient relating line emissivity with ionization parameter, assuming that locally the emission-line gas responds linearly to continuum variations  \citep{goa99a,kor04,goa12}.
The larger $\eta$ is, the more responsive the line is to continuum variations.
Although $\eta $ is model dependent, high-ionization emission lines generally have $\eta \sim 1$; Balmer lines have $\eta \sim 0.6$; and \ion{Mg}{2} emission line has $\eta \sim 0.2$ \citep{goa99a,kor04}.
We are able to interpret our result with respect to the weak variability of \ion{Mg}{2} as the consequence of the small responsivity $\eta$ for the low-ionization lines.

\subsection{\ion{Fe}{2} Pseudo-continuum Variability}
\label{sec:6_3}

The \ion{Fe}{2} emission line is also a member of the low-ionization lines.
In the same way as the \ion{Mg}{2} emission line, the weak variability of \ion{Fe}{2} can be the consequence of the small responsivity $\eta$.
However, the situation may be more complicated in regards to the \ion{Fe}{2}.

The origin of \ion{Fe}{2} emissions is still controversial \citep[e.g.,][]{don11}.
\cite{bal04} showed that \ion{Fe}{2} emissions from photoionized broad emission line region (BELR) cannot explain all the \ion{Fe}{2} emission features, and suggested that a large amount of \ion{Fe}{2} emission originates in either a separate collisionally ionized component or BELR with considerable velocity structure (microturbulence).
The collisionally ionized component would not vary in response to continuum variations, thus \ion{Fe}{2} pseudo-continuum variability would be weak as a whole.
On the other hand, several authors suggest that \ion{Fe}{2} is emitted in infalling photoionized gas \citep{hu08,fer09}, and several recent observational evidences support this picture  \citep[][and references therein]{bar13}. 
When this is the case, the small amplitude of \ion{Fe}{2} variability can be explained as the consequence of the different origin of \ion{Fe}{2}, (i.e., \ion{Fe}{2} emission does not come from usual BELR, so it cannot vary in the same way as other BELs). 
However, more detailed investigation of the \ion{Fe}{2} emission is beyond the scope of this paper. 

\subsection{Balmer Continuum Variability}
\label{sec:6_4}

Figure~\ref{fig:sbb_vandenberk} indicates that the Balmer continuum is a dominant variable component in SBB region.
We can expect that the Balmer continuum and Balmer emission lines are formed coherently by some photoionization processes in the broad line region, as we can see in the fourth global eigenspectrum presented by \cite{yip04}.
There is no wonder that the Balmer continuum is variable like the Balmer emission lines (e.g., H$\alpha$ and H$\beta$).
The variability of the Balmer continuum has been discussed observationally in  \cite{wam90} and \cite{mao93}, and theoretically in \cite{kor01}.
Our result is the first study clearly showing the variability of the Balmer continuum emission in large quasar samples.

\cite{wil05} showed that the ratio spectrum of the composite difference spectrum to the composite spectrum was depressed around 3000\AA \ \citep[see also][]{bia12}. Photometric variability using structure function formalization  \citep{van04,mac12} also shows the depression around 3000\AA.
This behavior can be interpreted by our result about the variability and non-variability of SBB components.
Because the \ion{Fe}{2} pseudo-continuum ($\sim$ 2200-3000\AA) and \ion{Mg}{2} emission line ($\sim$ 2800\AA) exist in the composite spectrum although not in the composite difference spectrum, the ratio is significantly depressed.
Moreover, as discussed in \cite{bia12}, the Balmer continuum should obey the intrinsic Baldwin effect in the same way as Balmer emission lines, so the ratio spectrum around the Balmer continuum region should be slightly depressed.

\subsection{Brief Summary of This Section}
\label{sec:6_5}

In summary, our conclusions in this section are:
\begin{enumerate}
\item We first decompose the SBB spectral region of the high resolution SDSS quasar composite spectrum \citep{van01} into a power-law continuum, Balmer continuum, \ion{Fe}{2}, and \ion{Mg}{2} following \cite{kur07}.
Then, we investigate the color$-$redshift relation of the composite spectrum with and without each component.
We confirm that the composite spectrum without \ion{Fe}{2} and \ion{Mg}{2} emission line components has similar color$-$redshift relations with the color of the flux difference spectrum.
This indicates that \ion{Fe}{2} and \ion{Mg}{2} have (in average) small variability, and the Balmer continuum is variable in broad spectral region.
This is the first time that the strong variability of Balmer continuum and small variability of \ion{Fe}{2} and \ion{Mg}{2} are confirmed as the statistical property of large quasar samples.
\item Small variability of the \ion{Mg}{2} emission line is naturally interpreted by the small responsivity $\eta$ of the low-ionization line to the ionizing continuum variation \citep{goa99a}.
\item Small variability of the \ion{Fe}{2} emission line is also interpreted by the small responsivity $\eta$ of the low-ionization line, but the origin of the \ion{Fe}{2} emission is controversial, so it is possible that the small variability of \ion{Fe}{2} is due to the different origin of \ion{Fe}{2} from normal photoionization BELR.
\item It is no wonder that the Balmer continuum is variable like the other Balmer emission lines because we can expect that they are formed coherently by some photoionization processes in the broad line region.
\end{enumerate}

\section{Conclusions}
\label{sec:7}

We have explored the wavelength dependence of quasar long term variability.
We have introduced the flux$-$flux correlation method, and applied it to a database of SDSS Stripe~82 multi-epoch five-band light curves for spectroscopically confirmed quasars.
This method enables us to infer the spectral variability without suffering from time-averaged baseline flux, which includes time-constant quasar host galaxy flux contamination and time-averaged emission line flux contaminations.

In this paper, we focused on the statistical properties of the quasar UV$-$optical variability, and
we conclude:
\begin{enumerate}
\item In Section~\ref{sec:4}, we show that the continuum component of the flux difference spectra is (on average) well approximated by a power-law shape with $\alpha_{\nu}^{\text{dif}}\sim +1/3$.
Then, we compare the $b(s,l)-$redshift relation with ``the standard disk with varying mass accretion rate model'' \citep{per06}.
We concluded that the flux difference spectrum is flatter (bluer) than the model prediction.
In particular, we confirm that the model-predicted ``UV turnover'' is not seen in the flux difference spectrum.
This is surprising because not only the standard accretion disk model \citep{sha73} but also any accretion disk models cannot produce such a flat spectrum as $\alpha_{\nu}^{\text{dif}}\sim +1/3$.
We stress the importance of the correction for Galactic extinction, without which the flux difference spectrum happens to have similar spectral shape with the model spectrum.

\item In Section~\ref{sec:5}, we identify the features due to, for instance, Balmer emission lines and Balmer continuum variability in the $b(s,l)-$redshift relation.
We show that some of the features seen in $b^{\text{ratio}}(s,l)-$redshift and $b(s,l)-$redshift relation are similar (e.g., the Balmer series emission), but they look significantly different around the SBB spectral region.
It indicates that the variability of \ion{Mg}{2} emission lines and \ion{Fe}{2} pseudo-continuum emission seem to be relatively weaker than other emission line components, such as the Balmer series emission lines.

\item In Section~\ref{sec:6}, we first decompose the SBB spectral region of the high resolution SDSS quasar composite spectrum \citep{van01} into a power-law continuum, Balmer continuum, \ion{Fe}{2}, and \ion{Mg}{2} following \cite{kur07}.
Then, we investigate the color$-$redshift relation of the composite spectrum with and without each component.
We confirm that the composite spectrum without \ion{Fe}{2} and \ion{Mg}{2} emission line components has similar color$-$redshift relations with the color of the flux difference spectrum.
This indicates that \ion{Fe}{2} and \ion{Mg}{2} have (in average) small variability, and the Balmer continuum is variable in broad spectral region.
This is the first time that the strong variability of the Balmer continuum and the small variability of \ion{Fe}{2} and \ion{Mg}{2} are confirmed as the statistical property of large quasar samples.
The small variability of the \ion{Mg}{2} emission lines and \ion{Fe}{2} emission lines is naturally interpreted by the small responsivity $\eta$ of the low-ionization line to the ionizing continuum variation \citep{goa99a}.
However, the origin of the \ion{Fe}{2} emission is controversial, so it is possible that the small variability of \ion{Fe}{2} is due to the different origin of \ion{Fe}{2} from normal photoionization BELR.
It is no wonder that the Balmer continuum is variable like the other Balmer emission lines, as we can expect that they are formed coherently by some photoionization processes in the broad line region.
\end{enumerate}

In the present work, we have limited our discussion to the UV$-$optical variability and UV$-$optical data.
However, it is clear that the AGN variability should be understood as multi-wavelength phenomena.
The correlation between the X-ray and UV$-$optical variability is   definitely important.
The X-ray reprocessing model, which also predicts the bluer-when-brighter UV$-$optical color variability with $\alpha^{\text{dif}}_{\nu}\sim+1/3$ if simple geometry for the X-ray emitting region and the accretion disk is assumed, may be the alternative model for AGN UV$-$optical variability \citep[``the lamppost model''; e.g.,][]{tom06,cac07,lir11,gil12,che13}.
This model also explains the variability time-scale (nearly simultaneous UV-to-optical variability) as the light-crossing time of the accretion disk.
However, the geometry around the accretion disk and the X-ray emitting region is not well known, and definitive conclusions about the validity of the X-ray reprocessing model have not been obtained yet.
The correlation between UV$-$optical and X-ray variability has been observed in several AGNs \citep[e.g.,][]{bre09,cam12}, but sometimes conflicting results with the X-ray reprocessing model have been obtained  \citep[e.g.,][]{mao02,she03,mar08}.
Future multi-wavelength intensive monitoring for more AGNs will be needed to clarify the origin of the X-ray$-$UV$-$optical correlation.

The present work deals with the properties of spectral variability only in the composite sense.
In a second paper, we would discuss the dependence of the spectral variability on several physical parameters, and (in)validity of the physical models for continuum variability in more detail.

\acknowledgments

We thank Makoto Kishimoto for useful discussions and comments.
This work is supported by Grants-in-Aid for Scientific Research (22540247, 25287062).
Data analysis were in part carried out on common use data analysis computer system (pc08, IDL8.1) at the Astronomy Data Center, ADC, of the National Astronomical Observatory of Japan.

Funding for the SDSS and SDSS-II has been provided by the Alfred P. Sloan Foundation, the Participating Institutions, the National Science Foundation, the U.S. Department of Energy, the National Aeronautics and Space Administration, the Japanese Monbukagakusho, the Max Planck Society, and the Higher Education Funding Council for England. The SDSS website is \href{http://www.sdss.org/}{http://www.sdss.org/}.

The SDSS is managed by the Astrophysical Research Consortium for the Participating Institutions, which are the American Museum of Natural History, Astrophysical Institute Potsdam, University of Basel, University of Cambridge, Case Western Reserve University, University of Chicago, Drexel University, Fermilab, the Institute for Advanced Study, the Japan Participation Group, Johns Hopkins University, the Joint Institute for Nuclear Astrophysics, the Kavli Institute for Particle Astrophysics and Cosmology, the Korean Scientist Group, the Chinese Academy of Sciences (LAMOST), Los Alamos National Laboratory, the Max-Planck-Institute for Astronomy (MPIA), the Max-Planck-Institute for Astrophysics (MPA), New Mexico State University, Ohio State University, University of Pittsburgh, University of Portsmouth, Princeton University, the United States Naval Observatory, and the University of Washington.

\appendix

\section{A: Model spectra used in Section~5}
\label{sec:A}

\begin{figure}[tbp]
 \begin{center}
  \includegraphics[clip, width=3.2in]{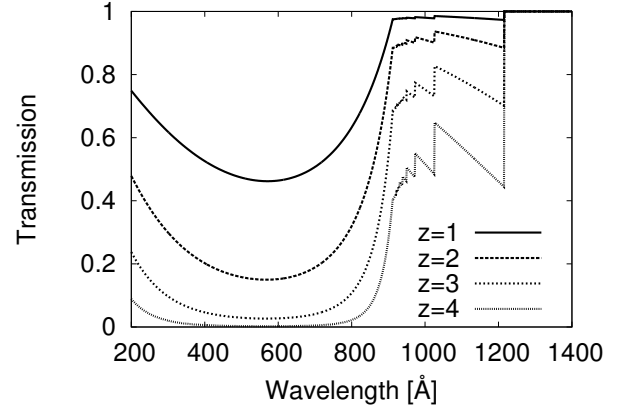}
  \vspace{-0.3cm}
 \end{center}
 \caption{Transmissivity $\exp(-\tau_{\text{eff}}(z,\lambda))$ as a function of rest-frame wavelength, where $\tau_{\text{eff}}$ is the effective optical depth of the IGM \citep{mei06}.}
 \label{fig:igm_transmissivity}
\end{figure}

Here we show how model spectra in Section~\ref{sec:5} are calculated.
\begin{description}
\item [Broad emission lines] $f^{\text{dif}}_{\nu}(\text{BEL})$ include H$\alpha$, H$\beta$, H$\gamma$, \ion{Mg}{2}, \ion{C}{3}],\ion{C}{4}, \ion{Si}{4}, Ly$\alpha$, which are the eight strongest BELs in the SDSS quasar composite spectrum \citep{van01}, and calculated as
\begin{equation}
f^{\text{dif}}_{\lambda}(\text{BEL})=R\frac{\sigma_{\lambda}(\text{Ly$\alpha$})}{\sigma_{\lambda}}\exp\left(-\left(\frac{(\lambda-\lambda_{\text{rest}})^2}{2\sigma_{\lambda}^2}\right)\right)
\end{equation}
where $\lambda_{\text{rest}}$ (rest-frame central wavelength for an emission  line), $R$ (relative strength for each BEL), and $\sigma_{\lambda}$ (line width) are given as Table \ref{composite_lines} for each emission line \citep{van01}.
\begin{deluxetable}{cccc}
\tablecaption{The Composite Spectral Property of Broad Emission Lines}
\tablehead{\colhead{Line} & \colhead{$\lambda_{\text{rest}}$[\AA]} & \colhead{$R$(Rel.Flux[100$\times$ $F$/$F$(Ly$\alpha$)])} & \colhead{$\sigma_{\lambda}$[\AA]}}
\startdata
H$\alpha $& 6564.61 & 30.832 & 47.39 \\ 
H$\beta  $& 4862.68 & 8.649  & 40.44 \\ 
H$\gamma $& 4341.68 & 2.616  & 20.32 \\ 
\ion{Mg}{2}& 2798.75 & 14.725 & 34.95 \\
\ion{C}{3}]& 1908.73 & 15.943 & 23.58 \\ 
\ion{C}{4}& 1549.06 & 25.291 & 14.33 \\ 
\ion{Si}{4}& 1396.76 & 8.916 & 12.50 \\ 
Ly$\alpha$& 1215.67 & 100.000 & 19.46
\enddata
\label{composite_lines}
\tablerefs{\cite{van01}.}
\end{deluxetable}
\item [Balmer and Paschen continuum emission]
For $f^{\text{dif}}_{\lambda}(\text{BaC})$ and $f^{\text{dif}}_{\lambda}(\text{PaC})$, we assume gas clouds of uniform temperature ($T_e$=15000~K) that are partially optically thick  \citep{die03,kur07}.
We assume that the Balmer continuum spectrum variability below the Balmer edge $\lambda_{\text{BE}}=3646$\AA \ is described by
\begin{equation}
f^{\text{dif}}_{\lambda}(\text{BaC})=R^{\text{BaC}}B_{\lambda}(T_e)\left( 1-\exp\left(-\tau_{\text{BE}}\left(\frac{\lambda}{\lambda_{\text{BE}}}\right)^3\right)\right)
\label{bac_cal}
\end{equation}
where $B_{\lambda}(T_e)$ is the Planck function at the electron temperature $T_e$, $\tau_{\text{BE}}$ is the optical depth at the Balmer edge $\lambda_{\text{BE}}=3646$\AA \ , and $R^{\text{BaC}}$ is the normalized flux density at the Balmer edge  \citep{gra82}.
The optical depth is fixed at $\tau_{\text{BE}}=1$  \citep{kur07} and  $R^{\text{BaC}}$ (with relative strength to other emission lines) is determined by the integrated flux ratio $F^{\text{BaC}}$/$F($H$\alpha)=1.80$  \citep{gra82}.
Note that in the actual quasar spectra, higher order Balmer lines are merging to a pseudo-continuum, yielding a smooth rise to the Balmer edge  \citep{wil85,die03}.
In the same way as the Balmer continuum, the Paschen continuum variability is expressed by
\begin{equation}
f^{\text{dif}}_{\lambda}(\text{PaC})=R^{\text{PaC}}B_{\lambda}(T_e)\left( 1-\exp\left(-\tau_{\text{PE}}\left(\frac{\lambda}{\lambda_{\text{PE}}}\right)^3\right)\right)
\end{equation}
where $\lambda_{\text{PE}}=8208$\AA \, the optical depth is fixed at $\tau_{\text{PE}}=0.25$, and $R^{\text{PaC}}$ is determined by the integrated flux ratio $F^{\text{PaC}}$/$F($H$\alpha)=1.17$ \citep{gra82}.
\item [\ion{Fe}{2}  pseudo-continuum emission] The ``SBB'' is composed not only of the Balmer continuum and \ion{Mg}{2} BEL, but also of \ion{Fe}{2} pseudo-continuum emission.
To model the \ion{Fe}{2} pseudo-continuum emission, we adopt a UV \ion{Fe}{2}  emission template presented by \cite{tsu06}.
As we do not know the relative strength of \ion{Fe}{2} emission to the power-law continuum or the other components in the SBB, we choose to scale the peak value of the \ion{Fe}{2} template at $\lambda=2481.9299$\AA\ to have $f_{\lambda}=0.3$, then
\begin{equation}
f^{\text{dif}}_{\lambda}(\text{\ion{Fe}{2}})=f_{\lambda}(\text{\ion{Fe}{2}}\text{\ template,\ scaled}).
\end{equation}
\item [Effects of Ly$\alpha$ forest (IGM attenuation)]
In addition to the emission lines variability, we show in Figures~\ref{fig:itiran} and \ref{fig:itiran_2} the wavelength-dependent effects by IGM attenuation.
At a wavelength range shorter than 1216\AA \ , observed flux is attenuated due to intervening absorption systems \citep[e.g.,][]{mad95}.
To describe the IGM attenuation at these wavelengths, we convolve a power-law continuum with the \cite{mei06} optical depth, which is
\begin{equation}
f^{\text{dif}}_{\lambda}(\text{Ly$\alpha$}\text{\ forest})=e^{-\tau _{\text{eff}}(z,\lambda)}f_{\lambda}^{\text{dif}}(\text{continuum})
\end{equation}
where (redshift and wavelength dependent) $\tau _{\text{eff}}(z,\lambda)$  is defined as the sum of the optical depth of resonant scattering by Lyman transitions and photoelectric absorption.
The transmissivity function is shown in Figure~\ref{fig:igm_transmissivity}.
\end{description}

\begin{figure*}[!t]
\centerline{
\hspace{-0.3cm}
\includegraphics[clip, width=7.5in]{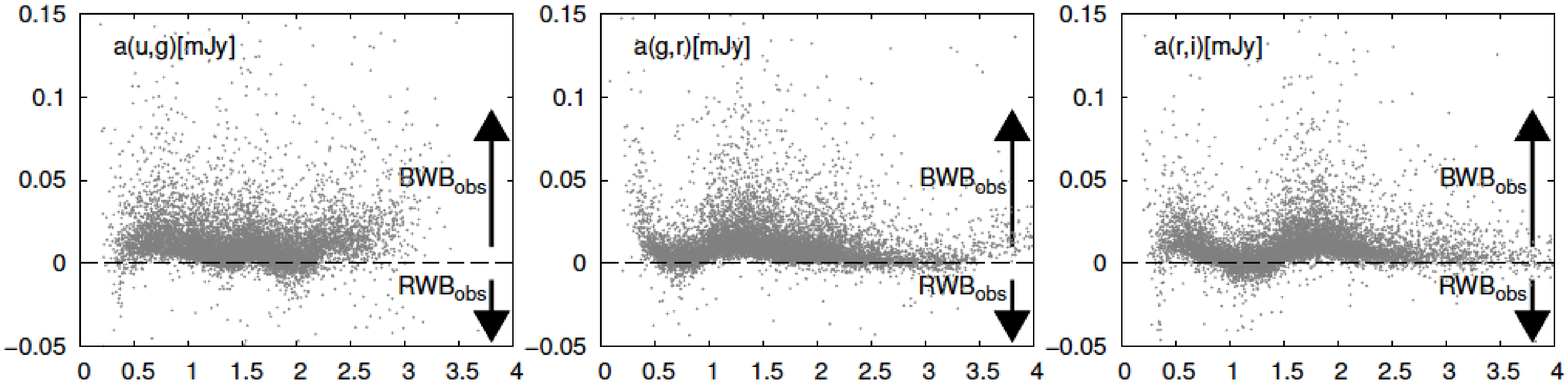}
}
\centerline{
\hspace{-0.2cm}
\includegraphics[clip, width=7.5in]{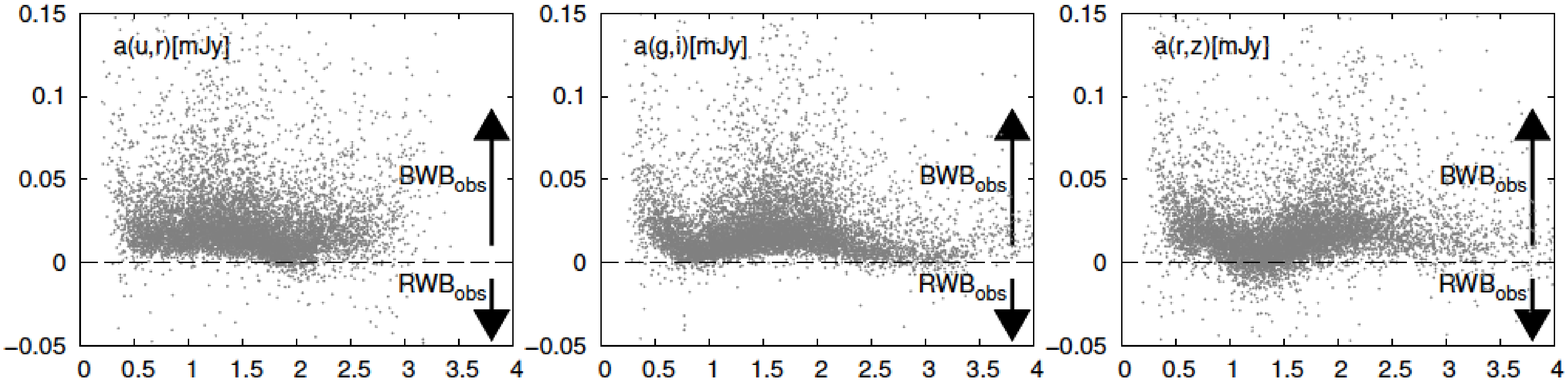}
}
\centerline{
\hspace{-0.2cm}
\includegraphics[clip, width=7.5in]{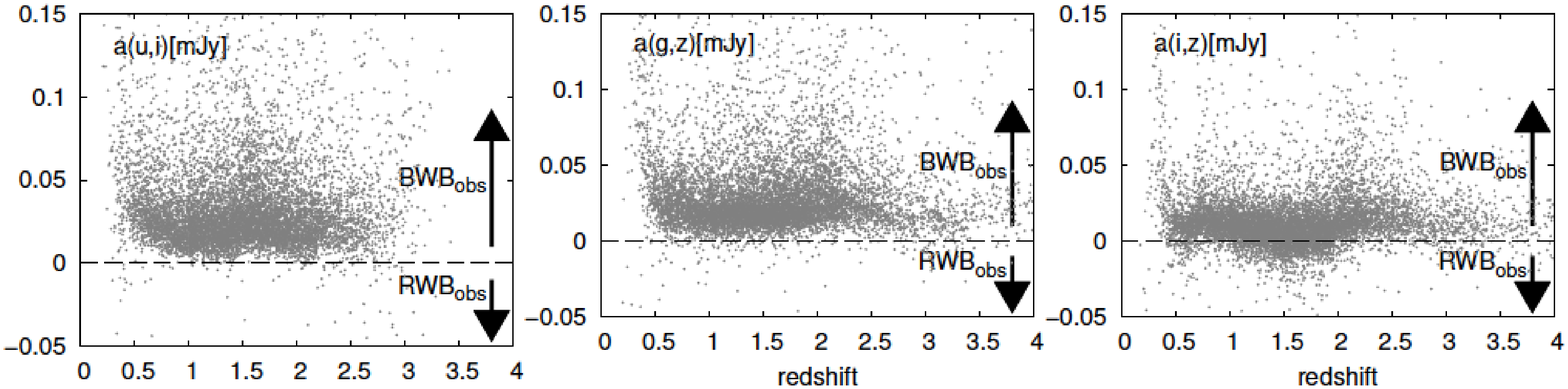}
}
\leftline{
\hspace{-0.67cm}
\includegraphics[clip, width=2.57in]{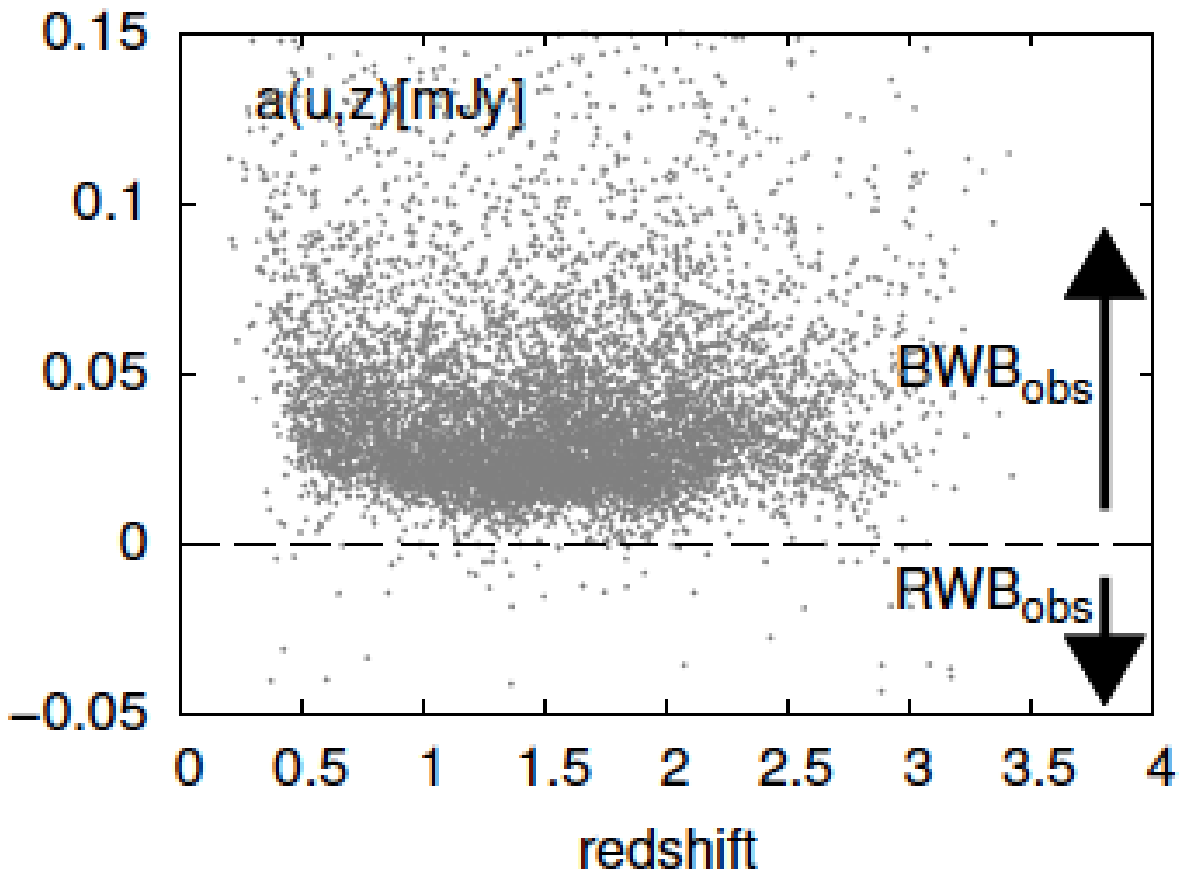}
}
\caption{Regression intercepts as a function of redshift (dots).
If $a(s,l)>0$, the ``observed'' color tends to become bluer when the quasar becomes brighter(BWB$_{\text{obs}}$ in Equation~(\ref{photometric_var})).
And if $a(s,l)<0$, the quasar has RWB$_{\text{obs}}$ trend.}
\label{fig:a_ugriz}
\end{figure*}

\section{B: Regression intercepts as a function of redshift}
\label{sec:B}

We here consider a flux$-$flux plot for a quasar light curve in which the regression line has positive intercept (as the case of Figure~\ref{fig:ff}).
$a(s,l)>0$ indicates the ratio of flux $f_{\nu}(l)/f_{\nu}(s)$, which is the slope of a straight line that passes through the origin and the observed data points at some epoch, becomes smaller when the quasar becomes brighter.
In this case, the photometric color $m_s-m_l$, defined as $+2.5\log (f_{\nu}(l)/f_{\nu}(s))$, becomes bluer-when-brighter.
Similarly, if the regression intercept is negative, the color of the quasar becomes redder-when-brighter  \citep{lyu93,hag97,sak10,sak11}.
Note that, as discussed in Section~\ref{sec:1}, the color variability trend referenced here is that of the ``observed'' photometric color, which does not directly mean the spectral hardening (softening) of the AGN intrinsic continuum emission.

Thus, the sign of the regression intercept in flux$-$flux space can be interpreted as the indicator of the ``observed'' photometric color variability.
We express this fact as 
\begin{eqnarray}
  \begin{cases}
    \text{BWB}_{\text{obs}} & (\text{if} \ a(s,l)>0) \\
    \text{RWB}_{\text{obs}} & (\text{if} \ a(s,l)<0)
  \end{cases}
\label{photometric_var}
\end{eqnarray}
where $\text{BWB}_{\text{obs}}$ and $\text{RWB}_{\text{obs}}$ indicates that the ``observed'' color becomes bluer when brighter and redder when brighter, respectively.
Figure~\ref{fig:a_ugriz} shows the regression intercepts as a function of redshift for each quasar.
Regression intercepts are generally positive, which indicates that the quasar color variability generally has $\text{BWB}_{\text{obs}}$ trend.
Note that this result is merely another expression of the result shown in Figure~\ref{fig:b_ugriz_ratio}, as the relation of $b^{\text{ratio}}(s,l)$ and $b(s,l)$ (Equation~(\ref{c_definition})) is actually determined by the sign of $a(s,l)$.

\clearpage

\bibliography{main}

\end{document}